\documentclass[preprint]{elsarticle}
\usepackage{multirow}
\usepackage{graphicx}
\usepackage{subcaption}
\usepackage{caption}
\usepackage{amsmath} 
\usepackage[percent]{overpic}
\usepackage{hyperref}

\begin{document}
\begin{frontmatter}
\title{\textbf{Performance evaluation of compact plastic scintillating fiber modules for muon tomography applications}}

\author[1,2]{Yiyue Li}
\author[2]{Huiling Li\corref{cor1}%
\fnref{fn1}}
\ead{huiling.li@iat.cn}
\author[2]{Hui Liang}
\author[2]{Cong Liu}
\author[1,2]{Chenghan Lv}
\author[2]{Hongbo Wang}
\author[1,2]{Weiwei Xu}

\affiliation[1]{organization={Institute of Advanced Technology, Shandong University},
addressline={17923 Jingshi Lu},
postcode={250100},
city={Jinan},
country={China}}
\affiliation[2]{organization={Particle Physics Research Center, Shandong Institute of Advanced Technology},
addressline={1501 Panlong Road},
postcode={250100},
city={Jinan},
country={China}}

\cortext[cor1]{Corresponding author}

\begin{abstract}
Muon tomography is a non-destructive imaging technique that exploits cosmic-ray muons from multiple directions. Its performance critically depends on the stability, active-area coverage, and spatial resolution of position-sensitive detectors. In this work, we report on the development of four compact scintillating fiber modules, each 100 cm long and composed of two staggered layers of 1 mm diameter fibers. The fibers are read out at one end by one-dimensional silicon photomultiplier arrays with a 2 mm pitch, coupled to Citiroc 1A-based front-end electronics. The modules were characterized with cosmic-ray muons, yielding a detection efficiency above 97\% and a mean spatial resolution of about 0.56 mm, with uniform response over different distances from the readout end. An imaging test of a lead block was also performed, and the reconstructed results are consistent with the block’s profile. These results demonstrate the suitability of this detector design for compact and large-area systems in muon tomography applications.
\end{abstract}

\end{frontmatter}

\section{Introduction}
Cosmic-ray muons are secondary particles naturally produced by the interactions of primary cosmic rays with the Earth’s atmosphere. At sea level, their flux is typically $\sim$$\rm{1\,cm^{-2}\,min^{-1}}$, with an average energy of $\sim$4 GeV~\cite{Navas:2024}. Owing to their strong penetrating power, Owing to their strong penetrating power, cosmic-ray muons enable muon tomography to non-destructively reconstruct the internal structures of large and dense objects by tracking these muons from multiple directions~\cite{Borozdin2003,Bonomi:2020,Tanaka:2023}. Depending on the physical principle, the technique can be classified into muon absorption tomography, which exploits the muon transmission rate, and muon scattering tomography, which relies on the measurement of muon deflection angles. In recent years, muon tomography has been successfully applied in diverse fields, including archaeology~\cite{Morishima:2017ghw}, geological exploration~\cite{Liu:2024}, nuclear waste monitoring~\cite{Mahon:2018}, and border security~\cite{Barnes:2023}.

Because of the relatively low flux of cosmic-ray muons and the complexity of practical environments, muon tomography systems demand detectors that combine high spatial resolution with large-area coverage and long-term operational stability. Plastic scintillator detectors are often preferred in such applications, as they can be flexibly manufactured in various geometries, maintain stable performance across different temperatures, offer high detection efficiency, and allow relatively low-cost large-area production. To achieve good spatial resolution, scintillators are usually cut into long bars with rectangular~\cite{Lesparre:2012} or triangular~\cite{Anstasio:2013,Wang:2024} cross sections, which are read out with wavelength-shifting fibers. New 3D printing techniques have also been developed to manufacture scintillators with more complex geometries~\cite{Yu:2025}. However, the spatial resolution of such scintillator detectors is typically at the millimeter level, and further improvement toward the sub-millimeter scale is difficult due to the constraints imposed by the bar shape and lateral size. 

The plastic scintillating fiber (SciFi) detector offers much finer granularity compared to bulk scintillator detectors, enabling superior spatial resolution. In addition, its long attenuation length and high light yield make it possible to construct SciFi detectors with lengths of several meters. These features render the SciFi detector a promising candidate for compact muon tomography systems. This potential has been demonstrated in Refs.~\cite{Clarkson:2014,Anbarjafari:2021} with round scintillating fibers, where grouped fibers were coupled with multi-anode PMTs or SiPM arrays to reduce the number of readout channels. To further enhance compactness and eliminate the laborious fiber-grouping process, we propose a SciFi detector employing a one-dimensional SiPM array readout~\cite{Chen:2023} combined with a potential SiPM multiplexing strategy~\cite{lv2025multiplexed} for sub-millimeter spatial resolution. In this work, we focus on the design and performance evaluation of SciFi modules with SiPM array readout, while the multiplexing strategy is discussed in a separate study.

The remainder of this paper is organized as follows. Section~\ref{sec:2} describes the design of the 100-cm-long SciFi modules aiming for sub-millimeter spatial resolution. Section~\ref{sec:3} presents the experimental setup and data acquisition system used for performance evaluation. Section~\ref{sec:4} details the data analysis procedures, including pedestal subtraction, gain calibration, signal clustering, tracking, and alignment. Section~\ref{sec:5} reports the experimental results on the detection efficiency and spatial resolution of the SciFi modules at different positions.  Section~\ref{sec:6} demonstrates a muon scattering tomography test and evaluates the imaging performance of the modules. Finally, Section~\ref{sec:Conclusion} concludes the work and provides an outlook for future developments.

\section{Construction of SciFi modules}\label{sec:2}

\begin{figure}[!t]
\begin{center}
\begin{tabular}{l}
\includegraphics[scale=0.43]{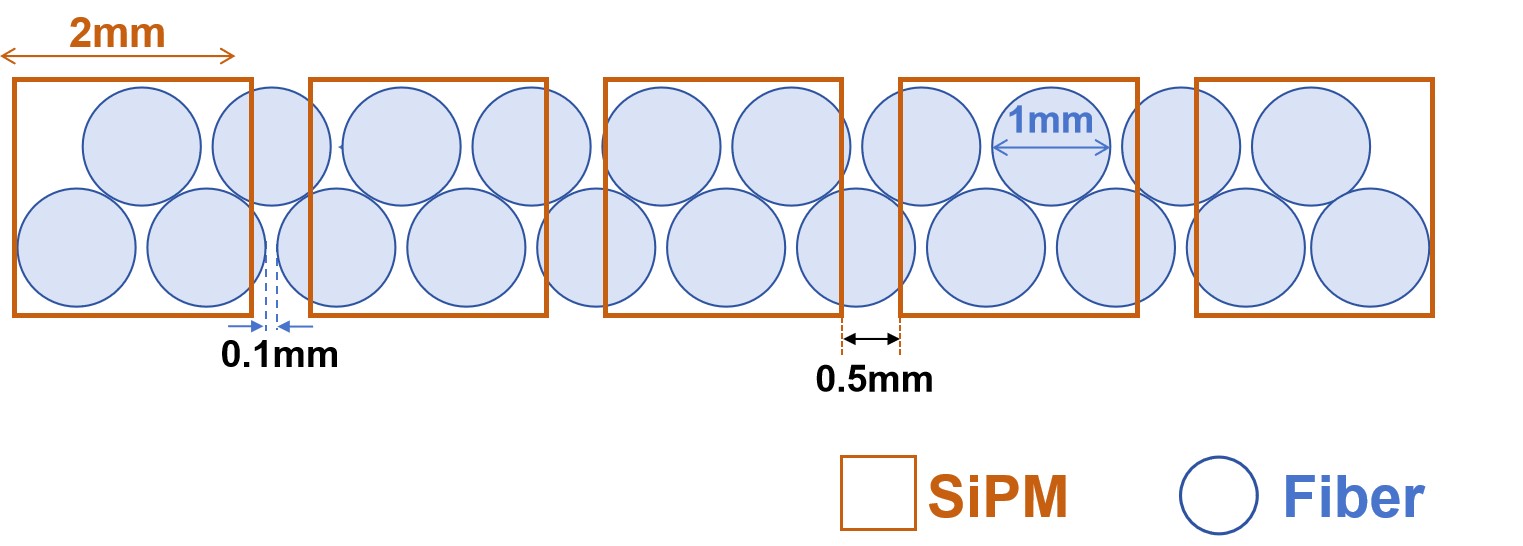}
\end{tabular}
\end{center}
\vspace{-0.8cm}
\caption{The design of one SciFi module with two staggered layers of 1 mm diameter fibers and one-dimensional SiPM array of 2 mm pitch size for sub-millimeter spatial resolution. }
\label{fig:SciFi_design}
\end{figure}
The spatial resolution of SciFi detectors with one-dimensional SiPM array readout is primarily determined by both the fiber diameter and the SiPM pitch. To achieve sub-millimetre precision, the preferred configuration would employ 1~mm diameter fibers coupled to SiPM arrays with a $1\times2$ $\rm{mm^{2}}$ or $1\times3$ $\rm{mm^{2}}$ active area per channel. However, such SiPM products are not yet commercially available. Therefore, as illustrated in Fig.~\ref{fig:SciFi_design}, the present SciFi modules are designed with two staggered layers of 1~mm diameter fibers, read out by SiPM arrays with a $2\times2$~$\rm{mm^{2}}$ active area per channel.
\begin{figure}
\begin{center}
\begin{tabular}{l}
\includegraphics[width=\linewidth]{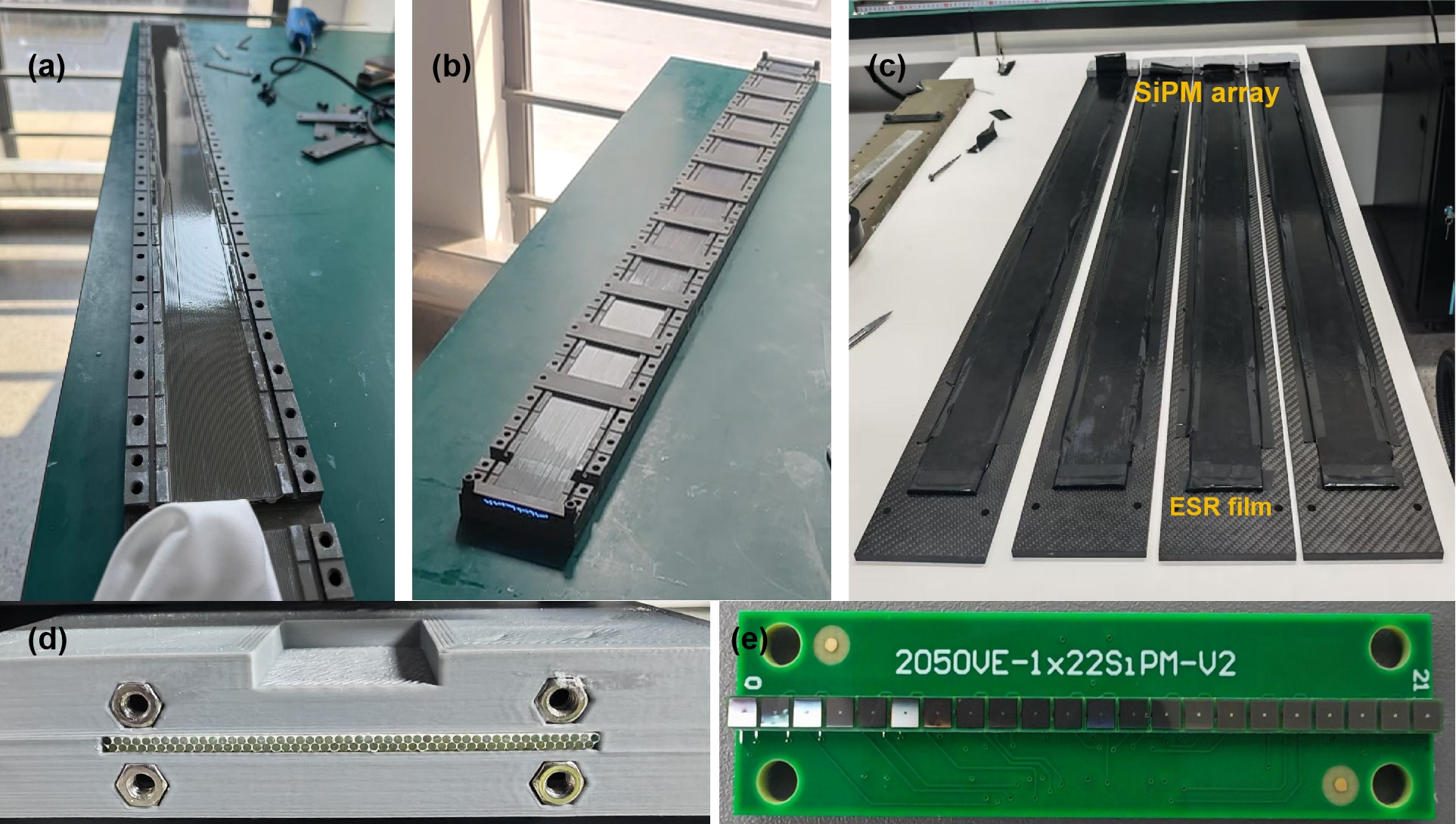}
\end{tabular}
\end{center}
\vspace{-0.6cm}
\caption{(a) Aluminium alloy mold with machined grooves, (b) first fiber layer placed along grooves of the mold, (c) the complete fiber modules with one end directly coupled to SiPM arrays and the other end covered by ESR film to reflect photons to the readout end, (d) the cross section of a SciFi module with two staggered layers of 1 mm diameter fibers, and (e) one-dimensional SiPM array composed of 22 Hamamatsu S13360-2050VE devices used at the readout end.}
\label{fig:SciFi_construction}
\end{figure}

Each compact SciFi module has dimensions of 100~cm (length)~$\times$~5.5~cm (width)~$\times$~2~mm (thickness) and is constructed using double-clad Kuraray SCSF-78M fibers~\cite{kurarayScintillatingFibers}. These fibers have a typical trapping efficiency of about 5.4\%, an attenuation length exceeding 4 m, and a light yield of roughly 8000 photons/MeV. Their emission spectrum spans 400–700 nm, peaking near 450 nm. To ensure a precise positioning throughout the module length, an aluminum alloy mold with precision-machined grooves is fabricated (Fig.~\ref{fig:SciFi_construction}(a)). After applying a release agent, the first fiber layer is placed in the grooves (Fig.~\ref{fig:SciFi_construction}(b)), followed by an adhesive mixed with $\rm TiO_{2}$ powder to reduce optical crosstalk between fibers. The second layer is then positioned in the interstitial grooves. The assembly is covered with Teflon sheets and is demolished once the adhesive solidifies. The fiber ends are cut and polished to ensure efficient optical coupling. Each module is mounted on a carbon-fiber support plate, with the readout end directly coupled to the SiPM arrays and the opposite end covered with an enhanced specular reflector (ESR) film via an air gap (Fig.~\ref{fig:SciFi_construction}(c)). In addition, four short SciFi modules (10~cm (length)~$\times$~5.5~cm~(width)~$\times$~2~mm (thickness)) are also produced using the same procedure but without ESR film to serve as muon trigger detectors in later tests.

To ensure efficient photon collection from the fibers, the dead area between adjacent SiPMs must be kept below 1 mm and minimized as much as possible. For this reason, Hamamatsu S13360-2050VE devices~\cite{HamamatsuS13360-2050VE} with a $\rm {2 \times 2}$ $\rm{mm^{2}}$ photosensitive area are selected to assemble the one-dimensional arrays as shown in Fig.~\ref{fig:SciFi_construction}(e). These devices employ through-silicon via (TSV) technology instead of conventional wire bonding, reducing the distance between the package edge and the photosensitive area to 0.2 mm on all four sides. An additional 0.1 mm margin was reserved to facilitate precise alignment and reliable soldering. The SiPMs are sensitive over the 320–900 nm wavelength range, matching well with the emission spectrum of the scintillating fibers. Their typical breakdown voltage is about 53 V and have peak photon detection efficiency reaches about 40\% at 3 V overvoltage around 450 nm. Furthermore, each channel has 50 $\rm \mu m$ pixel pitch and 1584 pixels, ensuring a linear response to photons emitted by the fibers when traversed by cosmic-ray muons.

\section{Experimental setup}\label{sec:3}

\begin{figure}[!h]
\centering
\hspace{-1cm}
\includegraphics[width=0.9\linewidth]{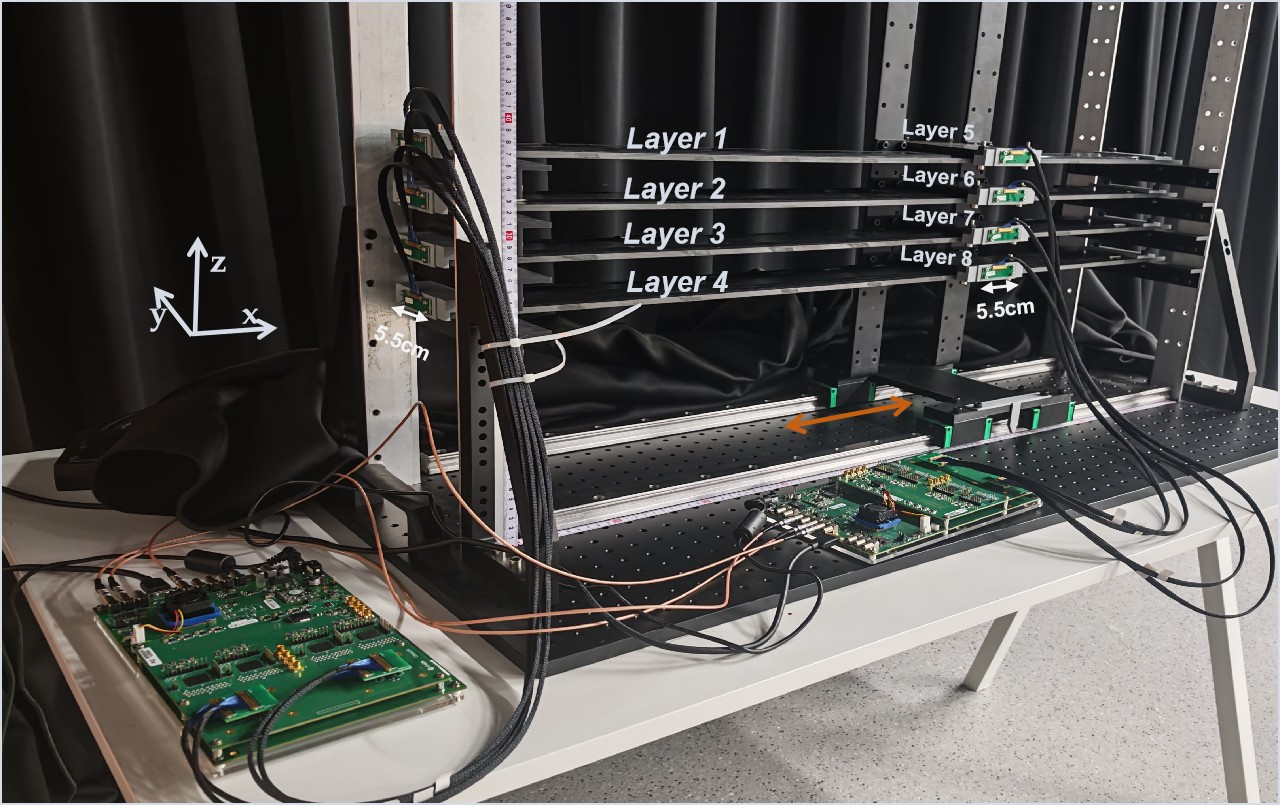}
\caption{Experimental setup of the SciFi modules with Citiroc1A-based readout electronics. The four short modules provide the muon trigger for the long SciFi modules during the cosmic-ray muon test. Their positions can be adjusted to record muon events at different locations along the long modules.}
\label{fig:scifi}
\end{figure}
The experimental setup is shown in Fig.~\ref{fig:scifi}. Four long SciFi modules (layers 1--4), each 100~cm in length and aligned along the $x$-axis, are arranged vertically with a spacing of approximately 5~cm along the $z$-axis. These layers are sensitive to the $y$-coordinate of incident particles. Four shorter modules (layers 5--8), each 10~cm in length and oriented along the $y$-axis, are placed perpendicularly to the long modules. The short modules can be translated as a group along the $x$-axis to scan muons striking different positions along the length of the long modules. The closest pair of short and long modules is separated by a vertical gap of about 0.5~mm, allowing for two-dimensional position measurements of muons. The long and short modules are read out by separate DT5550W data acquisition boards from CAEN, each equipped with four Citiroc 1A chips\cite{Citiroc1A} on a piggyback board and a 14-bit ADC. The SiPM array of each SciFi module is connected to a dedicated Citiroc 1A chip and biased at 57 V via the piggyback’s power supply module.

The Citiroc 1A is a 32-channel front-end ASIC optimized to amplify and integrate SiPM signals. Each channel includes two parallel preamplifiers, one in high-gain (HG) mode and the other in low-gain (LG) mode, extending the dynamic range up to about 2500 photoelectrons (p.e.). Each preamplifier output is processed by a slow shaper with an adjustable shaping time between 12.5 ns and 87.5 ns, followed by a peak-sensing circuit to measure the integrated charge. For triggering, the chip provides a 15~ns peaking-time fast shaper capable of discriminating signals down to 1/3 the signal amplitude of one p.e. in each channel, as well as an ASIC-level charge trigger formed by the logical OR of all 32 channels. The DT5550W board can then generate a global trigger by combining the logic OR outputs of its four Citiroc 1A chips.  

During the experimental tests of the SciFi modules, the HG signal was amplified by a factor of 26 and the LG signal by a factor of 2.6, with the slow shaping time of all channels fixed at 87.5 ns. Three dedicated data acquisition modes were implemented for pedestal estimation, gain calibration, and cosmic-ray muon detection, respectively:
\begin{itemize}
\item \textbf{Pedestal mode:}
SiPM signals were recorded without bias voltage using 1~kHz periodic triggers generated by the DT5550W software.

\item \textbf{Gain-calibration mode:}
In this mode, the two DT5550W boards operated independently. A global trigger was issued whenever the charge in any channel of the four Citiroc~1A chips exceeded about 5~p.e., prompting the readout of all SiPM channels of the four long (or short) modules by their respective DAQ boards. The threshold of 5~p.e. was empirically chosen to maintain a manageable trigger rate. Under these conditions, most recorded signals originated from SiPM dark noise or ambient radiation rather than cosmic-ray muons.

\item \textbf{Muon-test mode:}
For the cosmic-ray muon tests, the DT5550W board connected to the short modules was configured to generate ASIC-level charge triggers whenever any Citiroc~1A channel exceeded about 10~p.e. This threshold was chosen as a compromise between background suppression and muon detection efficiency. The trigger signal was then sent to the DT5550W board of the long modules as an external trigger. The total trigger latency, including cable delays, was measured to be approximately 40~ns using charge injection into a Citiroc~1A under identical acquisition conditions. This ensured that the trigger arrived prior to the peak-sensing stage of the long modules with 87.5~ns shaping time. Upon receiving the external trigger, the long-module data were read out. Complete events were subsequently constructed offline by matching the long- and short-module data according to their trigger IDs. Figure~\ref{fig:triggerRate} shows the hourly trigger rate of the experimental setup operated in muon-test mode, demonstrating stable and reliable performance over the full data-taking period.

\begin{figure}[!htbp]
\centering
\includegraphics[width=0.7\linewidth]{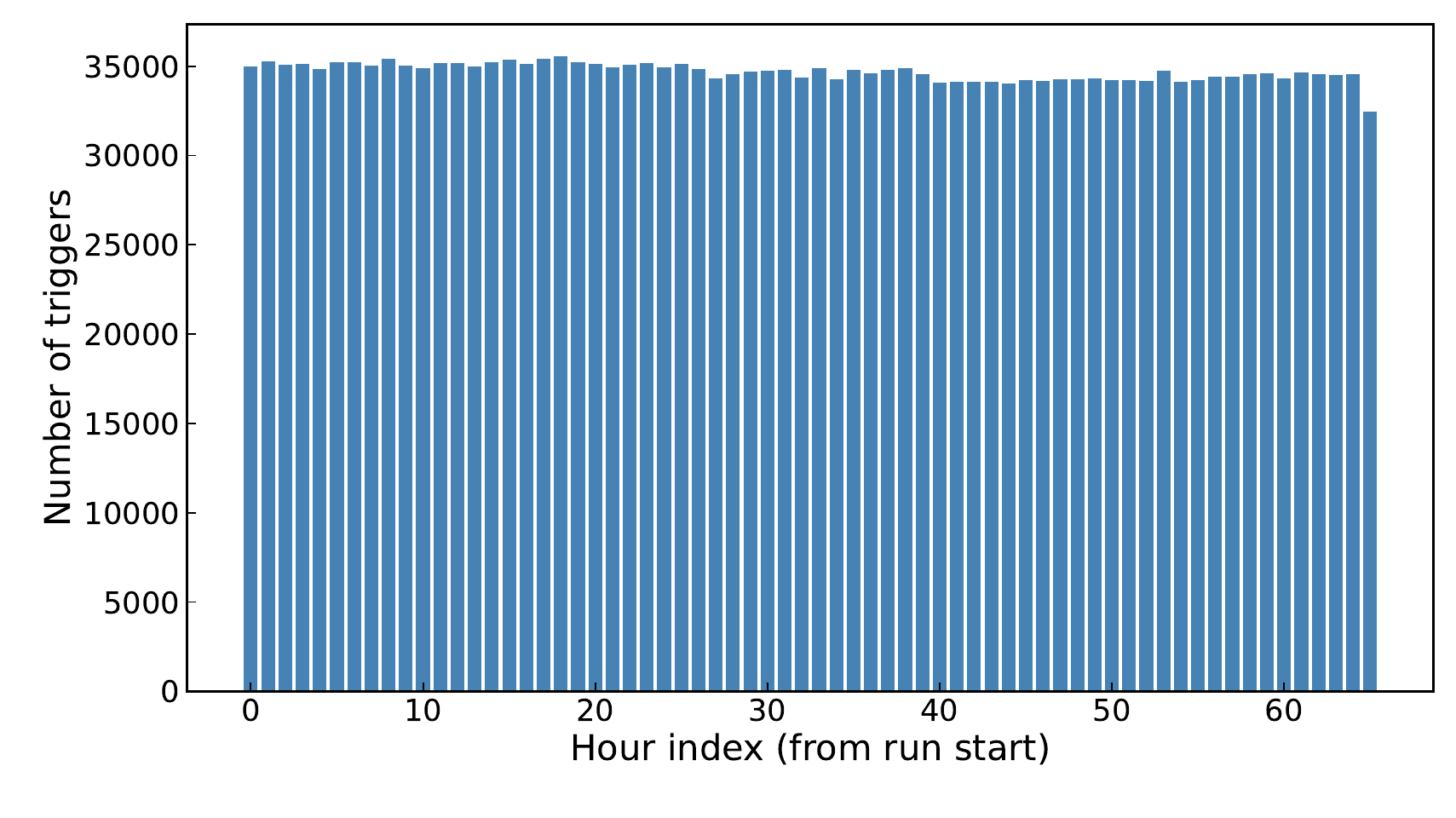}
\caption{Hourly trigger rate of the experimental setup operated in muon-test mode, demonstrating stable and reliable performance.}
\label{fig:triggerRate}
\end{figure}
\end{itemize}

\section{Data analysis}\label{sec:4}

\subsection{Channel calibration}
The raw signal from a fired SiPM is the sum of signals from all microcells that fired in Geiger mode. The number of fired microcells corresponds to the number of photoelectrons ($N_{\rm{pe}}$) for the channel, making $N_{\rm{pe}}$ proportional to the integral of the SiPM pulse, i.e., the total charge. Although all SiPM channels of a SciFi module were biased at the same voltage during the cosmic-ray muon test, variations in their breakdown voltages and the responses of different Citiroc1A channels resulted in non-uniform responses in $N_{\mathrm{pe}}$ among SiPM channels. To account for these variations, the pedestal must be subtracted from the raw data, and the gain of each SiPM channel must be calibrated individually.

\begin{figure}[!htbp]
\begin{center}
\begin{tabular}{l}
\hspace{-0.6cm}
\includegraphics[width=0.55\textwidth,height=4.8cm]{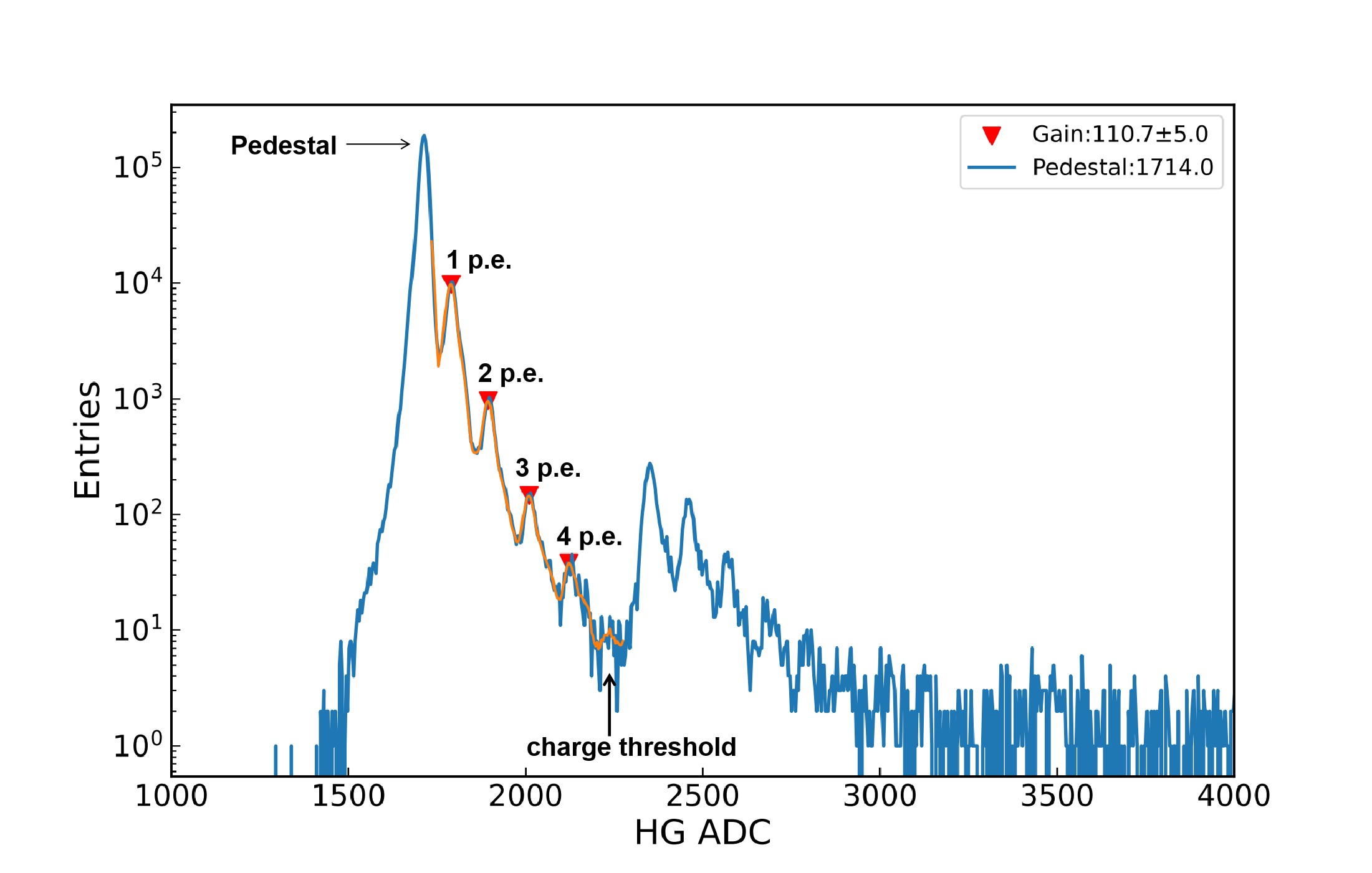}
\hspace{-0.6cm}
\includegraphics[width=0.55\textwidth,height=4.3cm]{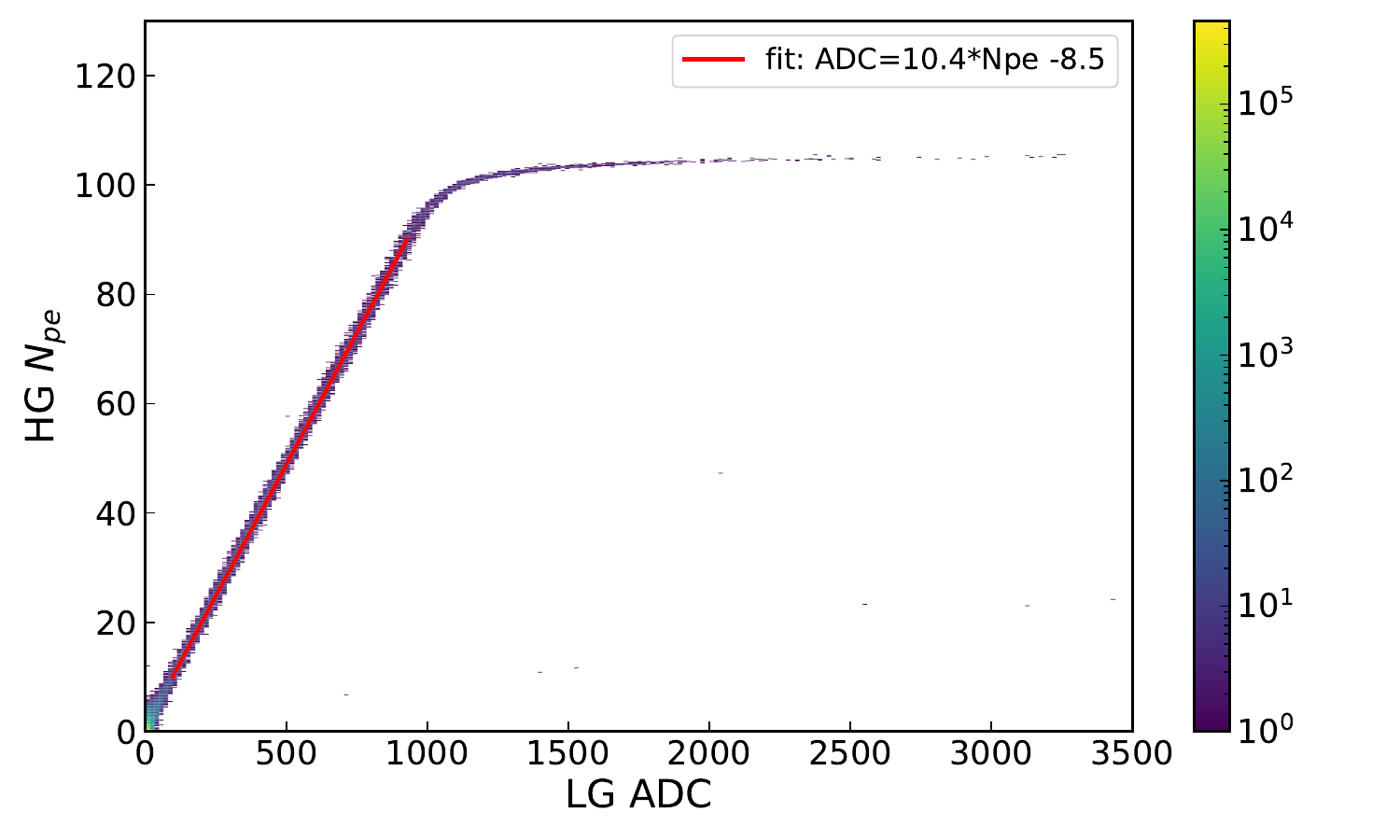}
\end{tabular}
\end{center}
\vspace{-0.6cm}
\caption{(Left) The raw ADC spectrum of one HG amplifier channel used to extract gain and pedestal values. (Right) LG signal after pedestal subtraction as a function of the calibrated HG $N_{\mathrm{pe}}$ for a representative SiPM channel, from which the gain in LG mode is obtained by a linear fit.}
\label{fig:gain}
\end{figure}

As shown in the left panel of Fig.~\ref{fig:gain}, the raw ADC spectrum of the HG amplifier signals obtained during gain calibration clearly exhibits an offset from zero. The pedestal is defined as the ADC value of the first and highest peak (from left to right), which is consistent with the value measured in pedestal mode with no high voltage applied to the SiPM arrays. The same procedure is applied to estimate the pedestal for the LG amplifier signals. The charge threshold set during gain calibration corresponds to the valley at approximately 5 p.e. The peaks between the pedestal and the threshold are identified, and the high gain, $G_{HG}$, is determined as the mean ADC separation between the adjacent peaks. As reported in Refs.~\cite{Impiombato:2020,Wu:2024}, the ADC difference between the pedestal and the 1 p.e. peak is systematically smaller than the actual gain, since the peak-sensing stage always samples the maximum signal value within the hold-delay window. Consequently, the 1 p.e. peak can be treated as a “fake pedestal” to estimate $N_{\rm{pe}}$ in an HG signal, as given by the equation:
\begin{equation}
N_{\mathrm{pe}} = \frac{ADC_{HG}-ADC_{1\mathrm{pe}}}{G_{HG}} + 1.
\end{equation}

\begin{figure}[h!]
\begin{center}
\begin{tabular}{l}
\hspace{-0.7cm}
\includegraphics[width=0.55\textwidth,height=5.5cm]{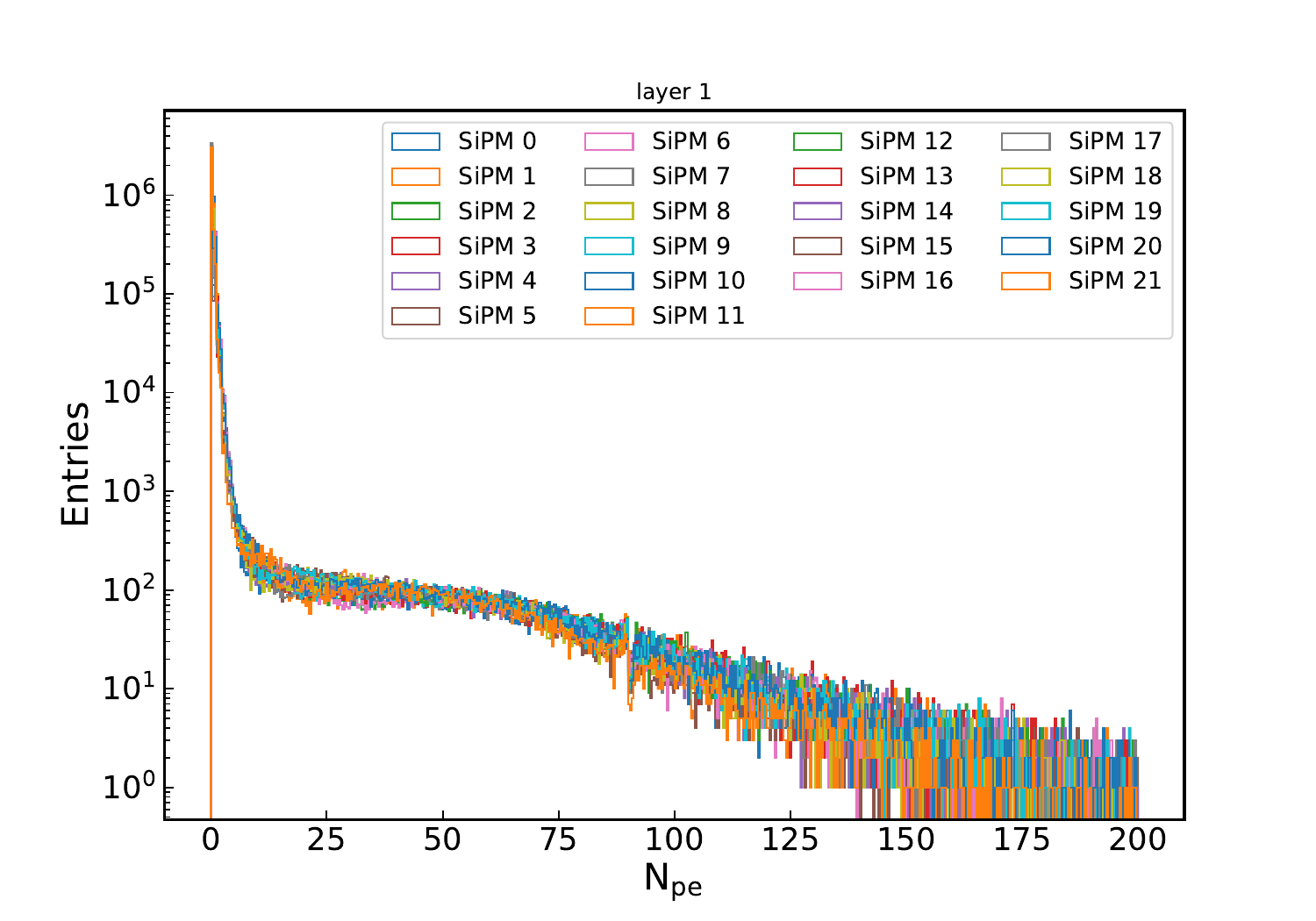}
\hspace{-0.6cm}
\includegraphics[width=0.5\textwidth,height=5cm]{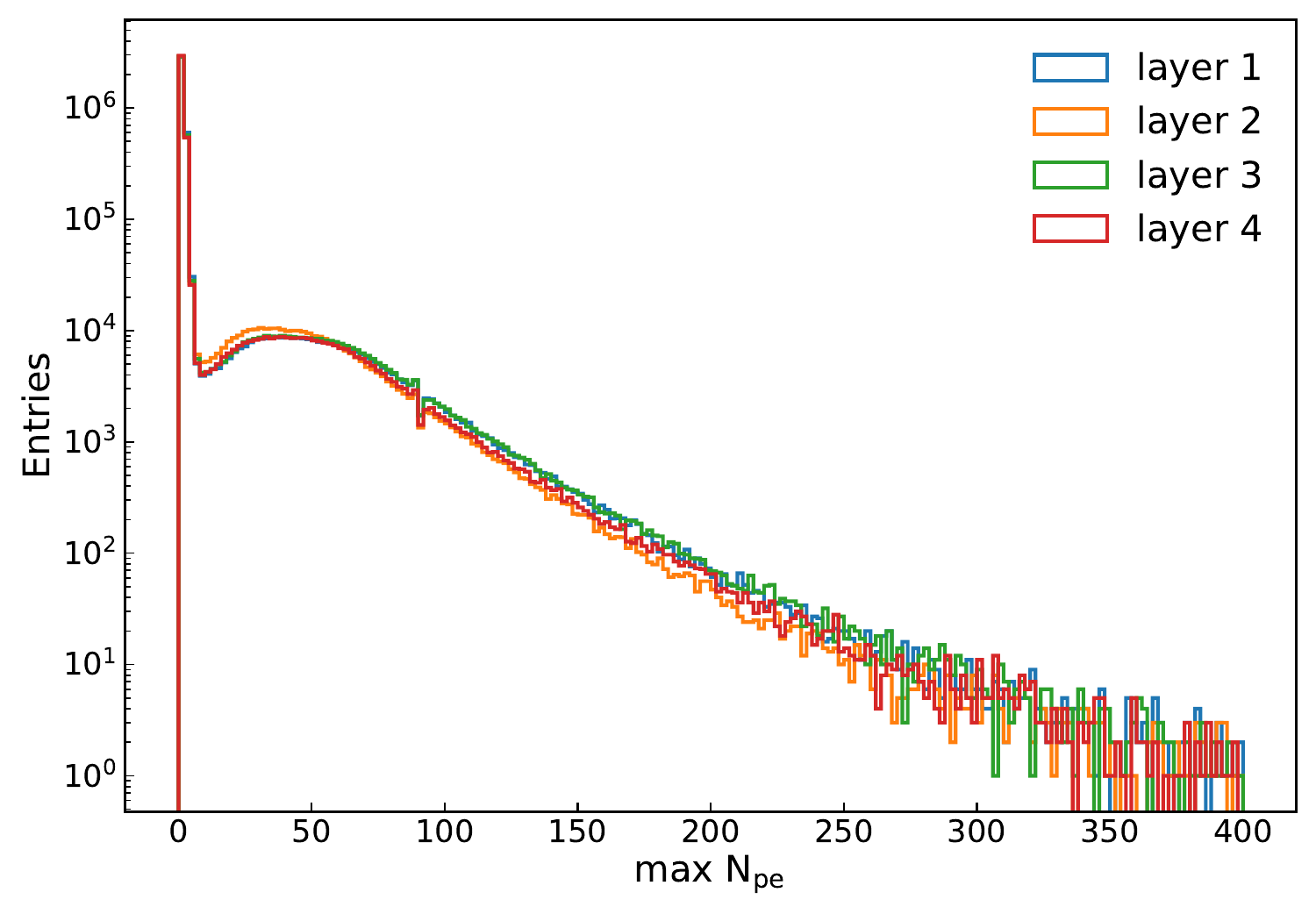}
\end{tabular}
\end{center}
\vspace{-0.6cm}
\caption{(Left) Calibrated SiPM signals from layer~1 acquired in muon-test mode after pedestal subtraction and gain calibration. (Right) Distribution of the SiPM channel with the maximum $N_{\mathrm{pe}}$ in each event for all four long modules when the short modules were centered at 52.5 cm from the readout end of the long modules.}
\label{fig:Npe}
\end{figure}

The gain in LG mode is inter-calibrated relative to the HG signals, as illustrated in the right panel of Fig.~\ref{fig:gain}. It is obtained as the slope of a linear fit using HG signals in the range of 10–90 p.e. The ratio of the obtained gains in HG and LG modes, as shown in Fig.~\ref{fig:gain}, is consistent with the amplifier value set in firmware. For larger SiPM signals ($>90$ p.e.), the HG response becomes non-linear. In such cases, LG signals are used to estimate the input $\rm{N_{pe}}$. After this calibration, the raw ADC data acquired in muon-test mode are converted into $N_{\mathrm{pe}}$ values. As shown in the left panel of Fig.~\ref{fig:Npe}, all 22 SiPM channels of layer 1 exhibit consistent signal spectra after pedestal subtraction and gain calibration. Moreover, the distribution of the SiPM channel with the maximum $N_{\mathrm{pe}}$ in each event is consistent across all four long modules, as illustrated in the right panel of Fig.~\ref{fig:Npe}. The slight deviation observed in layer 2 is attributed to differences in the procedure used when attaching the ESR films.

\subsection{Signal clustering and track reconstruction}\label{sec:cluster}
Cosmic-ray muons passing through a SciFi module may deposit energy in multiple fibers, thereby inducing SiPM signals in one or more neighboring channels. A group of such neighboring channels is defined as a \textit{cluster}. To select muon signals while suppressing background events from ambient radiation, the following procedure is applied: the seeding channel is identified as the one with a signal exceeding 8 p.e., corresponding to the valley in the $N_{\mathrm{pe}}$ distribution shown in the right panel of Fig.~\ref{fig:Npe}. Neighboring channels with signals above 6 p.e. are then included together with the seeding channel to form a cluster. The number of SiPM channels contained in a cluster is referred to as the cluster size, $cls$. The hit position of the cluster on the SciFi module is determined using the center-of-gravity (CoG) method:
\begin{equation}
x_{\mathrm{cog}} = \frac{\sum_{i} N_{i} x_{i}}{\sum_{i} N_{i}} ,
\end{equation}
where $x_{i}$ is the center position of the $i$th SiPM in the cluster, and $N_{i}$ is its corresponding signal size.

\begin{figure}[!htbp]
\centering
\includegraphics[width=0.8\linewidth]{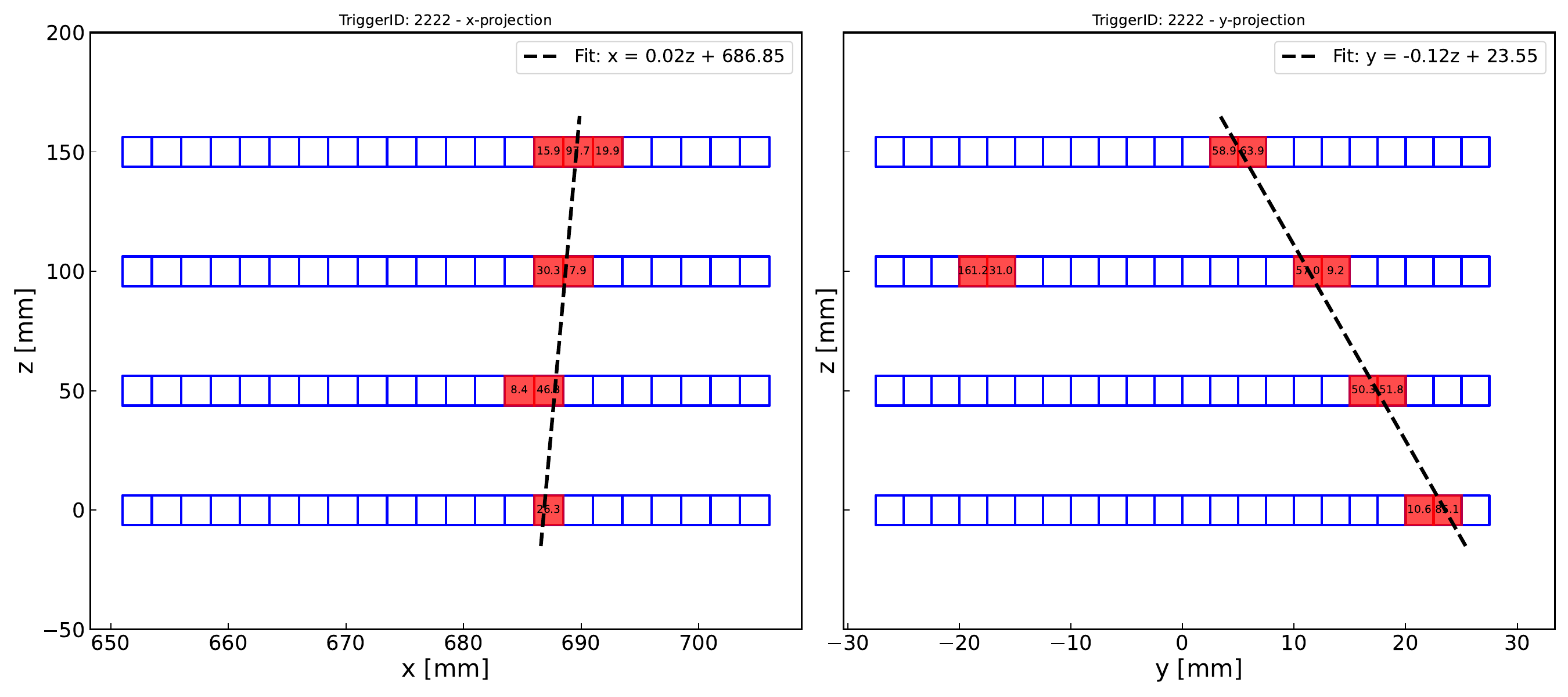} \\
\caption{Event display of reconstructed tracks in the $xz$ and $yz$ projection planes. The blue rectangles represent SiPM arrays, while the red rectangles indicate the signal cluster with black numbers as their $N_{\mathrm{pe}}$ values in each channel.}
\label{fig:trk_cls}
\end{figure}

\begin{figure}[!htbp]
\centering
\hspace{-0.1cm}
\includegraphics[width=0.5\linewidth]{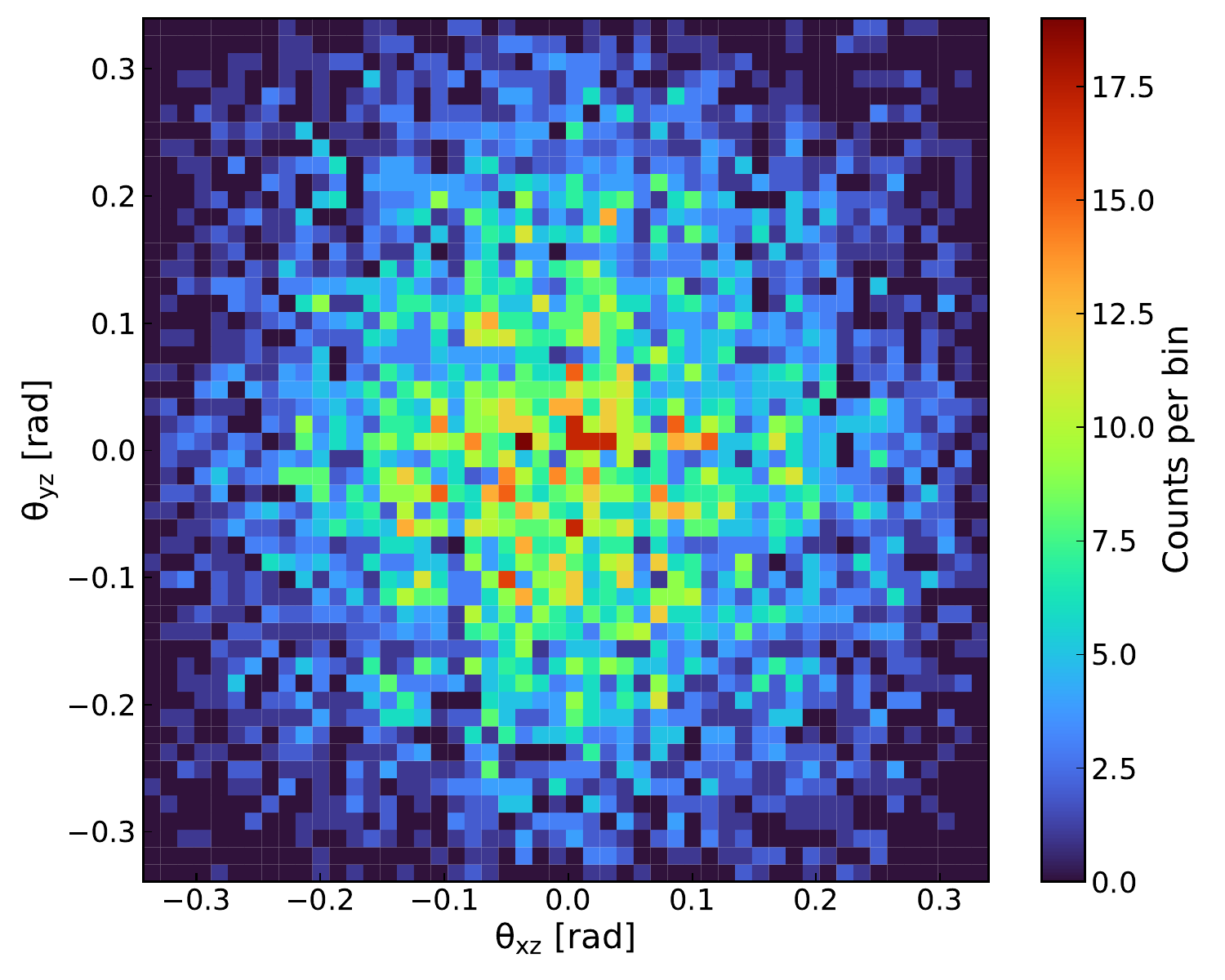} \\
\caption{The angular map of reconstructed tracks after alignment. Here the $yz$ tracks are reconstructed with layers 1,3,4, and the $xz$ tracks are reconstructed with layers 5,7,8.}
\label{fig:angular}
\end{figure}

Typically, muon events may result in signals from one or two channels in the SciFi module. Crosstalk between adjacent fibers due to imperfect glue coverage may also induce signals from $cls=3$, which was observed during quality inspections of the modules. To suppress background clusters or clusters contaminated by background channels, only those with a cluster size $cls \leq 3$ in a given layer are retained for track reconstruction. Particle tracks are reconstructed separately in the $xz$ and $yz$ planes. All possible combinations of clusters across different layers are scanned and tested with a straight-line fit using the least squares method, and the set of clusters yielding the smallest fitting $\chi^{2}$ is selected as the track candidate for that event. Fig.~\ref{fig:trk_cls} illustrates an event display of reconstructed tracks in the $xz$ and $yz$ projection planes after alignment, as explained in Sec.~\ref{sec:align}. Fig.~\ref{fig:angular} shows the two-dimensional angular distribution of reconstructed tracks after alignment to illustrate the field of view of the entire setup.

\subsection{Module alignment}\label{sec:align}
The fitting residual of a SciFi module, defined as the difference between the reconstructed cluster position and the extrapolated position from the fitted track onto the same layer, originates from the intrinsic detector resolution, multiple-scattering effects, and mechanical tolerances of the setup assembly. To obtain a more precise determination of the intrinsic position resolution, as discussed in Sec.~\ref{sec:posres}, a track-based alignment procedure is applied to all modules in the experimental setup.

In this procedure, a transverse translation $s$ and a small rotation angle $\theta$ around the $z$-axis are considered as free alignment parameters for each SciFi module. The alignment constants are determined by minimizing the global sum of squared residuals,
\begin{equation}
R^{2} = \sum_{N}\sum_{i}\left[ \left(y_{i}^{c}-y_{i}^{f}\right)^{2} + \left(x_{i}^{c}-x_{i}^{f}\right)^{2} \right],
\label{eq:alignment}
\end{equation}
where $N$ is the number of tracks, $x_{i}^{c}$ and $y_{i}^{c}$ are the corrected cluster positions of the nearest pair of short and long modules after applying the alignment parameters, and $x_{i}^{f}$ and $y_{i}^{f}$ are the extrapolated positions of the tracks fitted by corrected cluster positions in the $xz$ and $yz$ planes, respectively. The corrections are expressed as
\begin{equation}
\begin{aligned}
y_{i}^{c} &= y_{i}^{m} + x_{i}^{m}\cdot\theta_{y_{i}} + s_{y_{i}}, \\[6pt]
x_{i}^{c} &= x_{i}^{m} - y_{i}^{m}\cdot\theta_{x_{i}} + s_{x_{i}},
\end{aligned}
\end{equation}
where $x_{i}^{m}$ and $y_{i}^{m}$ denote the measured cluster positions in the paired short and long modules, respectively. 
\begin{figure}[!htbp]
\centering
\hspace{-0.1cm}
\includegraphics[width=\linewidth]{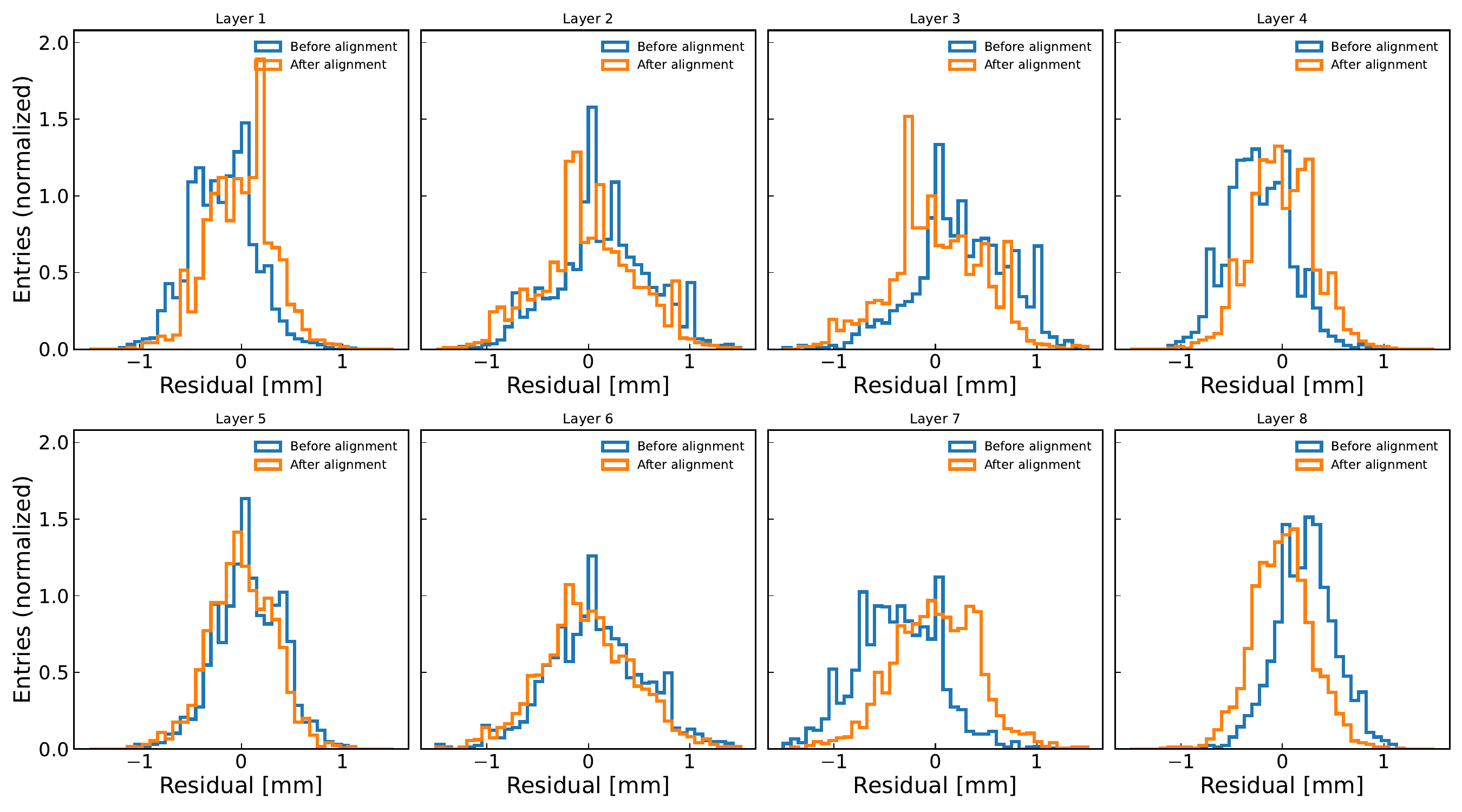}
\caption{The residual distributions of all SciFi modules before and after alignment.}
\label{fig:alignment}
\end{figure}

During the alignment procedure, only the cluster with the maximum signal on each layer is used in track fitting. Tracks are required to satisfy the following criteria: a cluster size of $cls \leq 3$ on all four layers, and a fit quality of $\chi^{2}<100$. The tracks reconstructed with cluster of $cls=1$ on all four long or short modules are dismissed to avoid biased residuals as explained in Sec.~\ref{sec:posres}. The minimization of Eq.~\ref{eq:alignment} is performed iteratively with the Gauss–Newton method~\cite{Blobel2007}. For each test position of SciFi modules, at least 1000 selected tracks are used for the alignment. Convergence to level of 0.1 mm of translation parameters is realized typically after 3 iterations. Fig.~\ref{fig:alignment} shows an example of the residual distributions with tracks of $\chi^{2}<10$ of all SciFi modules before and after alignment using muon test data, clearly demonstrating that the mean of the residuals shifts to zero after alignment.

\section{Module performance}\label{sec:5}

The detection efficiency and spatial resolution are key performance parameters of the SciFi modules, as they determine both the imaging time and the reconstruction accuracy in muon tomography. In this section, we take a single long SciFi module as the Device Under Test (DUT) and evaluate these parameters for each DUT individually. Measurements are performed at five positions along the $x$ axis, defined as the distance between the center of the short module and the readout end of the long module. These positions are realized by sliding the short modules within the setup. For each position, data are collected over periods longer than five days in muon-test mode.

\subsection{Detection efficiency}\label{sec:eff}
\begin{figure}[!htbp]
\centering
\includegraphics[width=0.5\linewidth]{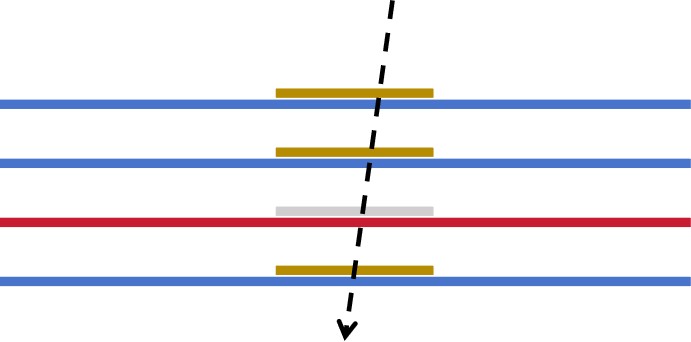}
\caption{Illustration of the detection efficiency calculation for a single long SciFi module (shown in red), with muon tracks selected using the short modules (shown in yellow) and the other long modules (shown in blue). In this example, layer 2 in red is treated as the DUT.}
\label{fig:cut}
\end{figure}

To determine the detection efficiency of the DUT, we define $\rm{N_{good}}$ as the number of muon tracks expected to pass through the fiducial volume of the DUT, and $\rm{N_{det}}$ as the number of these tracks that produce a valid cluster on the DUT. Fig.~\ref{fig:cut} illustrates an example that the detection efficiency calculation for layer 2 as the DUT, where the short modules except for the one paired with the DUT are employed to constrain the muon impact position, while the other long modules are used to reconstruct the reference tracks for efficiency determination. 

To be explicit, the selection of good muon events follows these criteria:
\begin{itemize}
\item For the short modules (excluding the one paired with the DUT), $xz$-plane tracks are reconstructed with a goodness-of-fit requirement of $\chi^{2}<2$. Clusters used in the reconstruction must have a signal of $N_{\mathrm{pe}}>8$~p.e. and a cluster size of $cls \leq 3$ in each participating layer. These conditions ensure that the selected muons traverse the intended measurement region on the long modules.
\item For the long modules (excluding the DUT), $yz$-plane tracks are reconstructed with $\chi^{2}<2$, and only clusters with $N_{\mathrm{pe}}>8$~p.e. and $cls \leq 3$ are accepted in each layer. Furthermore, the extrapolated track position on the DUT is required to fall within its active sensitive area.
\end{itemize}

\begin{figure}[!htbp]
\centering
\includegraphics[width=0.8\linewidth]{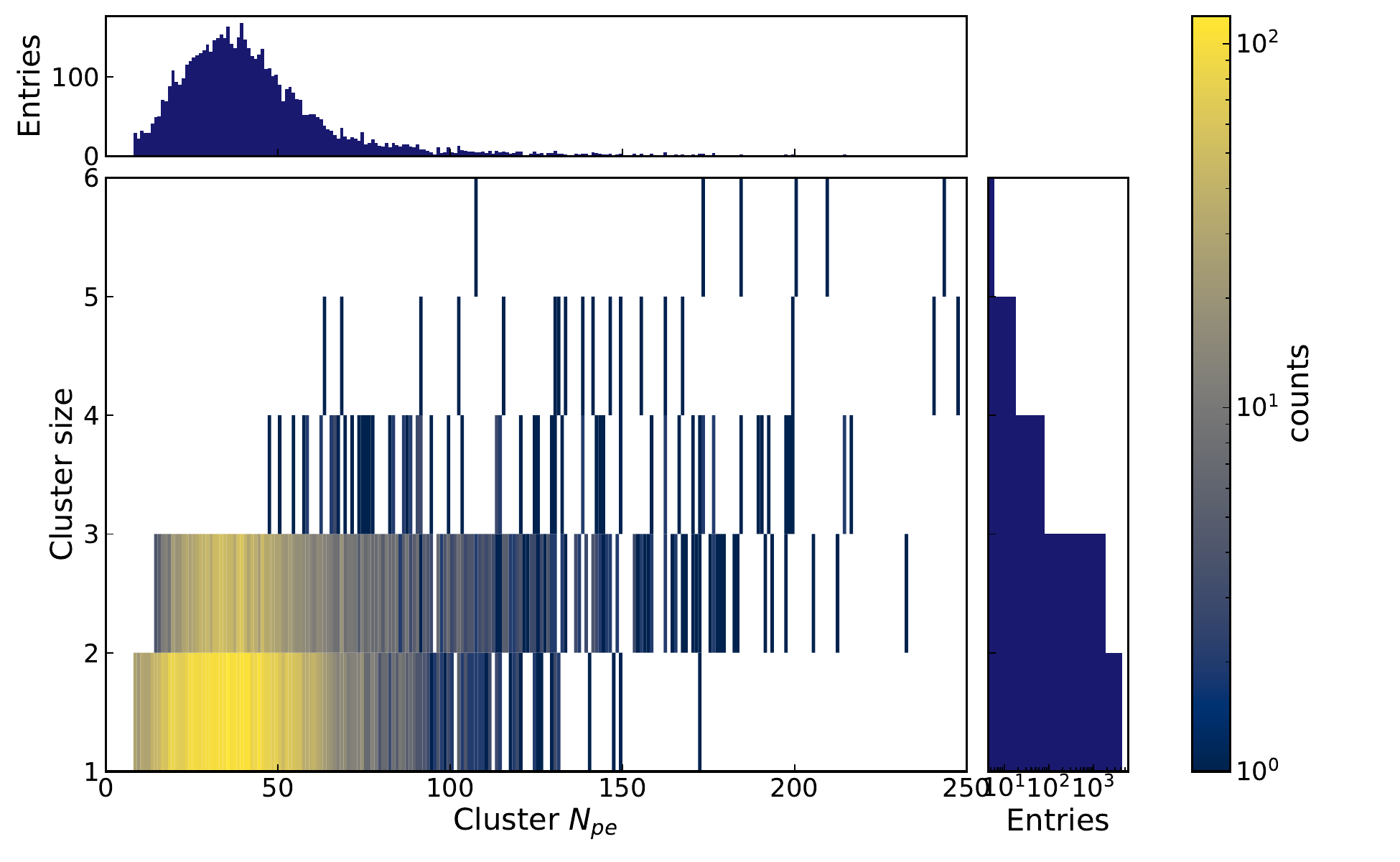}
\vspace{-0.3cm}
\caption{Example distribution of cluster size versus cluster signal $N_{\mathrm{pe}}$ for good events in layer~2, measured at a position of 52.5~cm.}
\label{fig:clusters}
\end{figure}
With these criteria, about 1000 good events per day are obtained for each DUT. Fig.~\ref{fig:clusters} shows, as an example, the cluster size and cluster signal distributions for the DUT in layer~2 at a measurement position of 52.5~cm. Most clusters have $cls \leq 2$, and the most probable value of $N_{\mathrm{pe}}$ is approximately 36 p.e., consistent with the module design expectation. Clusters with larger sizes are typically associated with higher signals, which can originate from muon hit combined with secondary particles produced in muon interactions, such as $\delta$ rays, or with ambient radiation.

For each good event, the DUT is considered to have successfully detected the muon if a cluster with $cls \leq 3$ is present on the DUT layer and the residual between the cluster position and the extrapolated track position in the $yz$ plane is less than 5~mm. Events with $cls=3$ are retained to maximize statistics while only marginally affects the position resolution. The residual threshold is particularly relevant for the outermost modules due to geometrical constraints. The detection efficiency is then defined as
\begin{equation}
\epsilon = \frac{N_{\mathrm{DUT}}}{N_{\mathrm{good}}}\; , 
\end{equation}
where $N_{\mathrm{DUT}}$ is the number of events with a valid cluster on the DUT and $N_{\mathrm{good}}$ is the total number of good events. 

\begin{figure}[!htbp]
\centering
\includegraphics[width=0.8\linewidth]{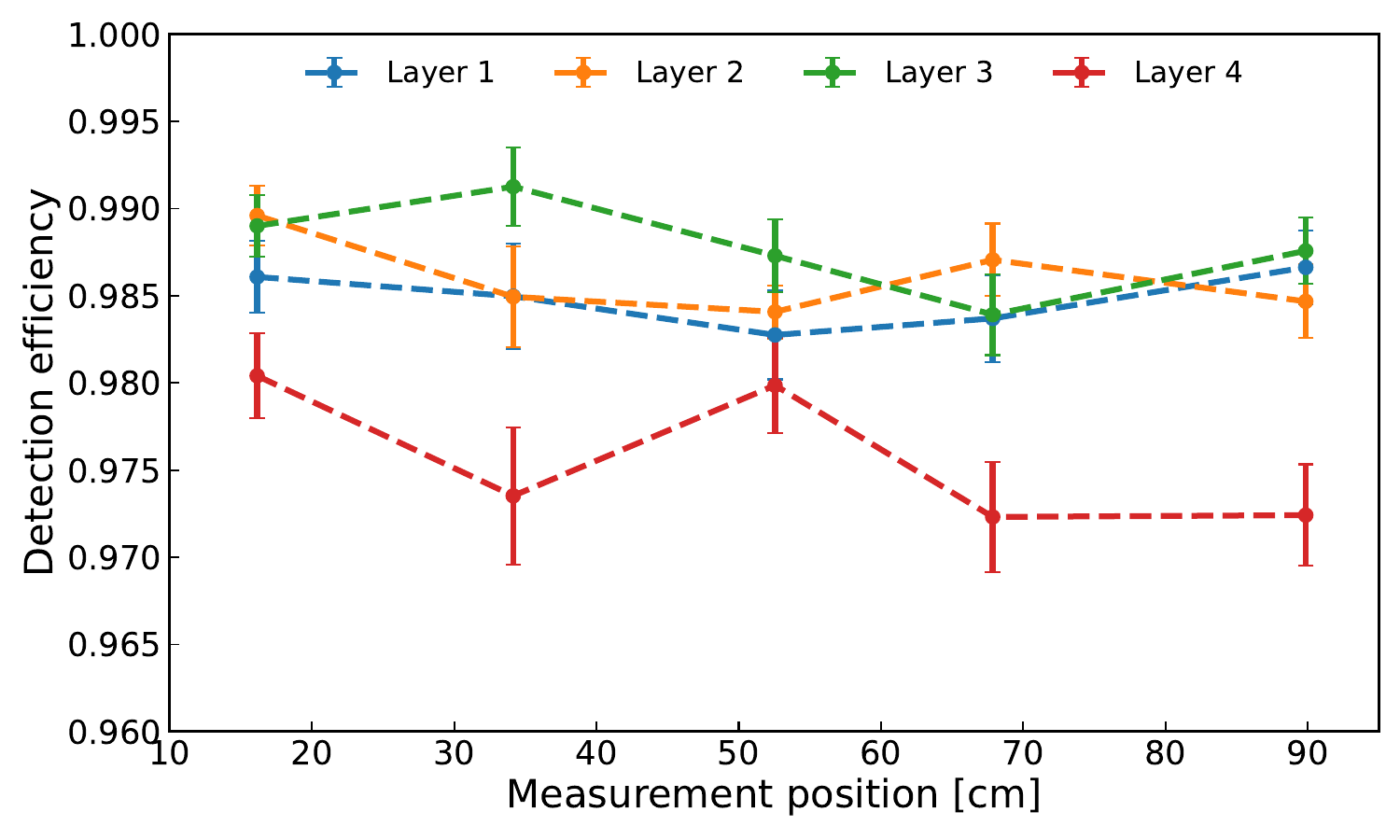}
\caption{Detection efficiency of the four long SciFi modules measured at different positions along their length. Error bars represent statistical uncertainties.}
\label{fig:detEff}
\end{figure}

Fig.~\ref{fig:detEff} shows the measured detection efficiencies of the four long SciFi modules at different measurement positions. The efficiencies are uniform along the module length and consistently exceed 97\%. To further verify that this measurement reflects the absolute detection efficiency, we varied the cluster signal threshold used to select good events (8, 15, 20, and 25~p.e.). The resulting detection efficiencies remain stable across these thresholds, confirming that the efficiency evaluation is robust.

\subsection{Spatial resolution}\label{sec:posres}
The spatial resolution of the long SciFi modules is primarily determined by the fiber diameter, the SiPM pitch, and the threshold settings used in the cluster-finding procedure. To qualitatively assess their influence, a preliminary Monte Carlo simulation of a single SciFi module was performed using the Geant4 toolkit~\cite{GEANT4:2002zbu}. Cosmic-ray muons were generated within the same zenith-angle range as in the experimental setup, following the modified Gaisser formula reported in Ref.~\cite{guan2015parametrization}. Detailed detector effects including SiPM dark noise, optical crosstalk between adjacent fibers, and the air gap between the fiber end faces and the SiPM array were not included.

\begin{figure}[!htbp]
  \centering
  \begin{subfigure}[t]{0.48\textwidth}
    \centering
    \begin{overpic}[width=\linewidth,height=3.8cm]{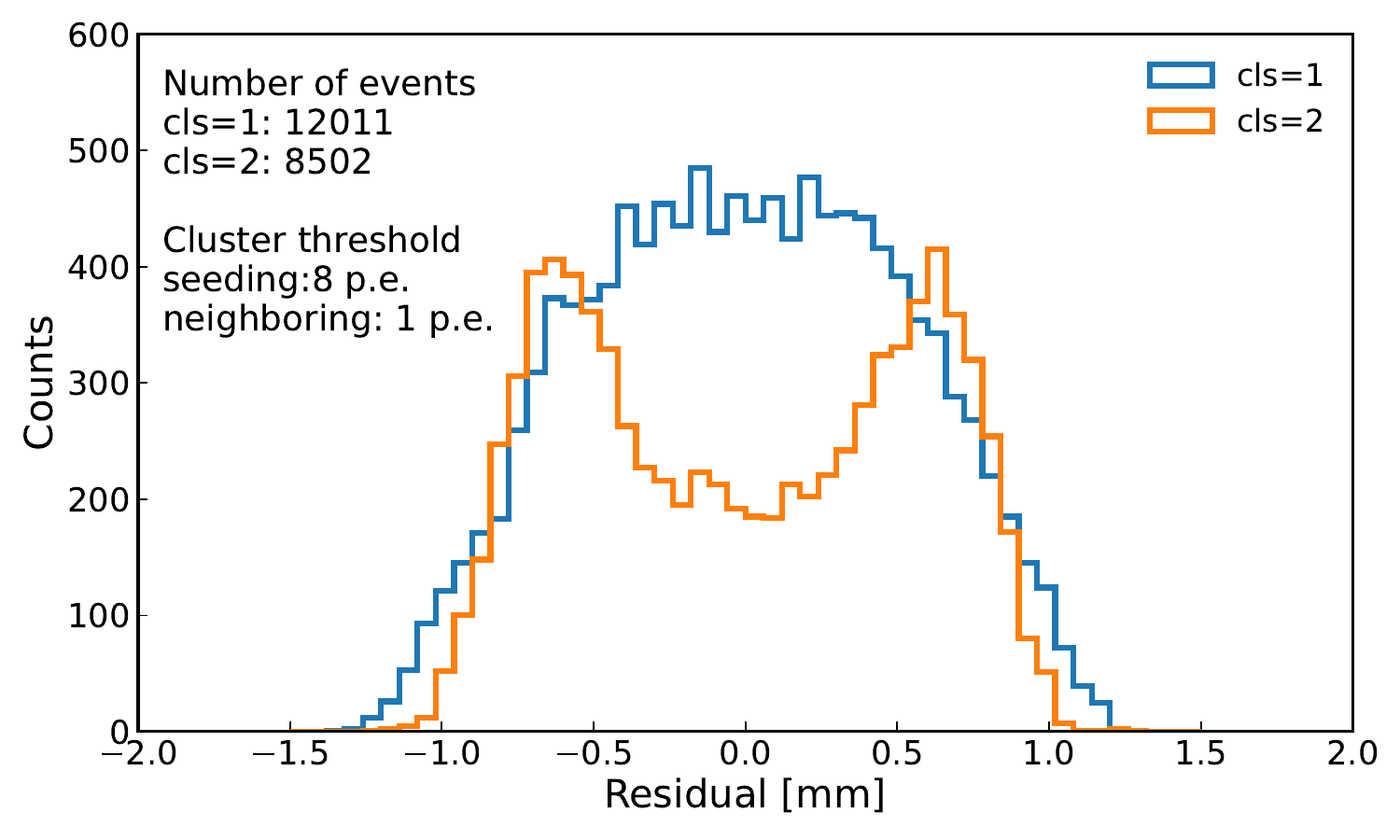} 
      \put(12,10){\small\textbf{(a)}}
    \end{overpic}
  \end{subfigure}
  \hfill
  \begin{subfigure}[t]{0.48\textwidth}
    \centering
    \begin{overpic}[width=\linewidth,height=3.8cm]{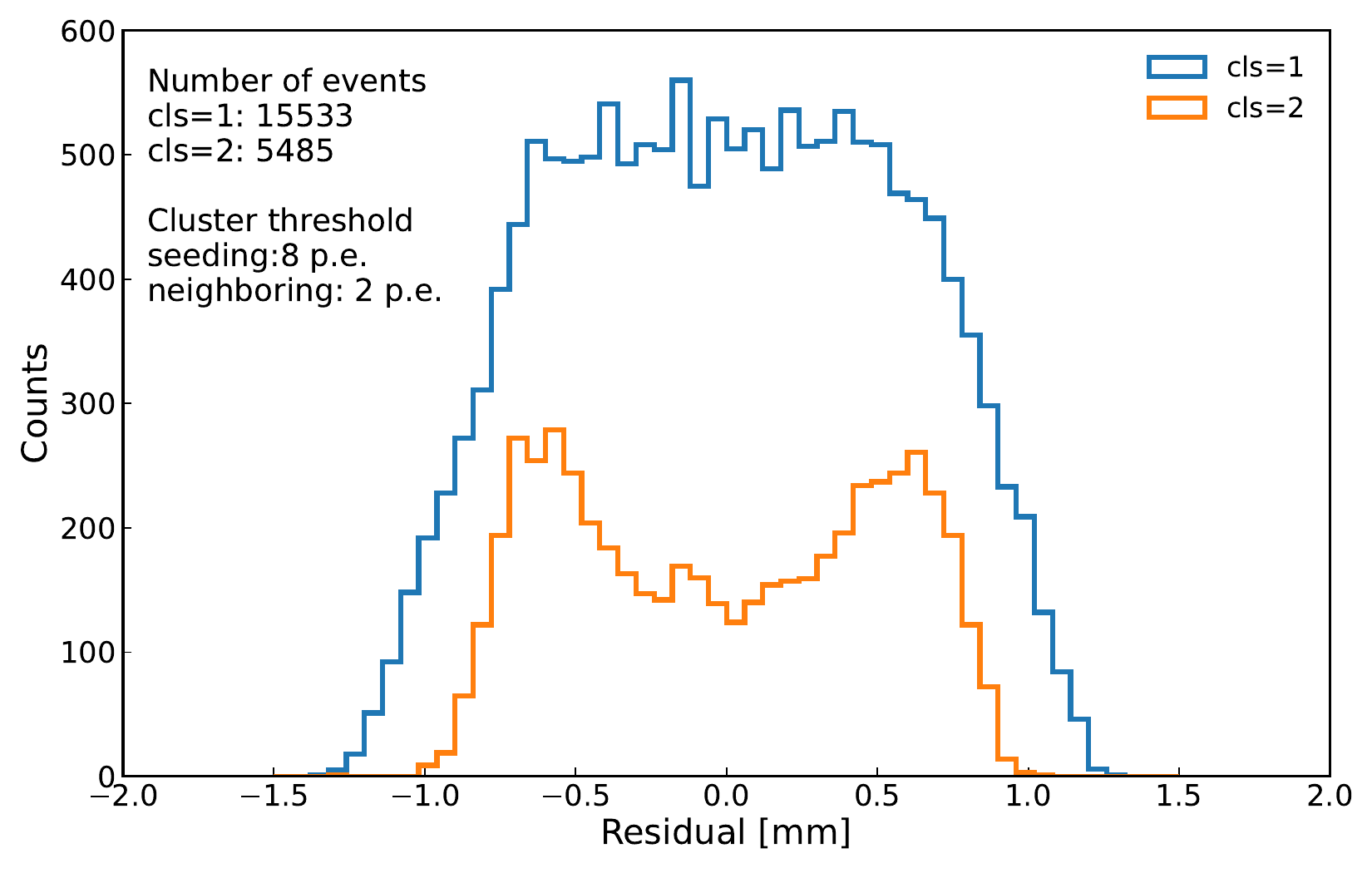}
      \put(12,10){\small\textbf{(b)}}
    \end{overpic}
  \end{subfigure}
  \hfill
  \\
  \begin{subfigure}[t]{0.48\textwidth}
    \centering
    \begin{overpic}[width=\linewidth,height=3.8cm]{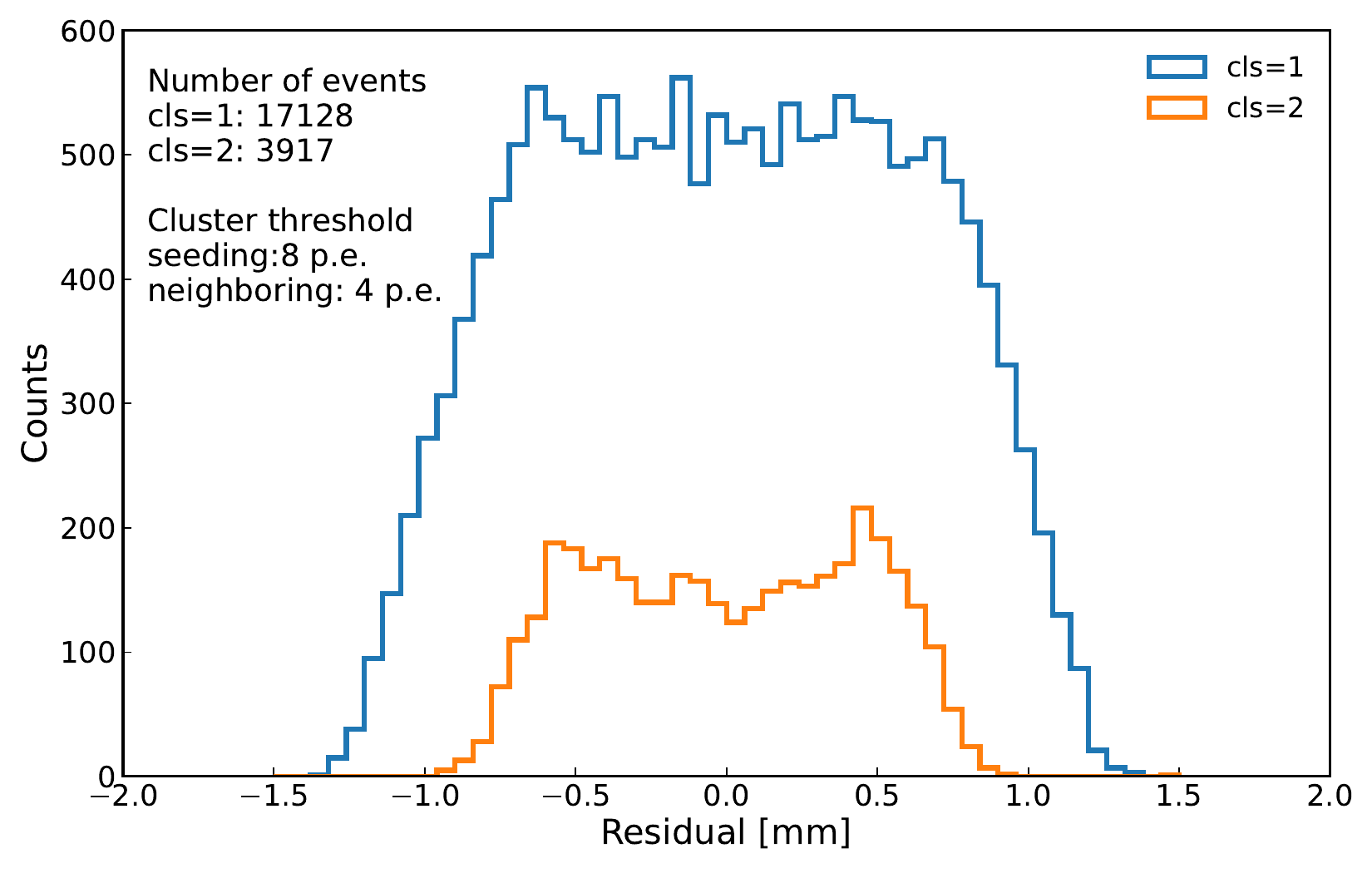}
      \put(12,10){\small\textbf{(c)}}
    \end{overpic}
  \end{subfigure}
  \hfill
  \begin{subfigure}[t]{0.48\textwidth}
    \centering
    \begin{overpic}[width=\linewidth,height=3.8cm]{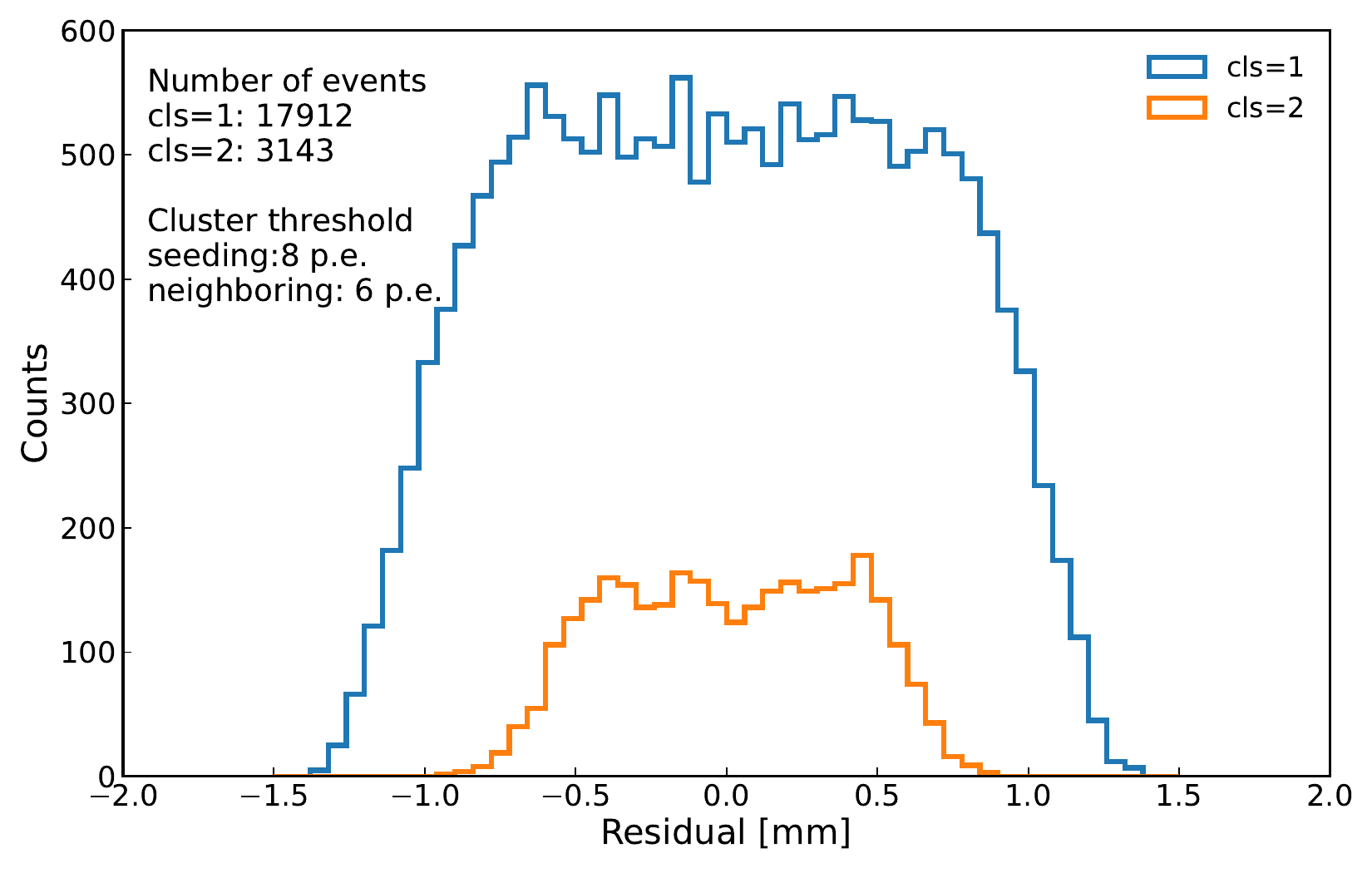}
      \put(12,10){\small\textbf{(d)}}
    \end{overpic}
  \end{subfigure}
  \caption{Residual distributions of simulated events with $cls=1$ and $cls=2$ by using different threshold values during cluster finding.}
  \label{fig:residual_cls_MC}
\end{figure}

The same cluster-finding algorithm described in Sec.~\ref{sec:cluster} was applied, and the simulated cluster signal distributions were tuned to reproduce the most probable values observed in Fig.~\ref{fig:clusters} by adjusting the light yield of the scintillating fibers. For each event, the residual between the reconstructed cluster position and the true hit position was evaluated under four different cluster-threshold configurations, as shown in Fig.~\ref{fig:residual_cls_MC}. Both the full width at half maximum (FWHM) of the residual distribution and the fractions of events with $cls=1$ and $cls=2$ exhibit dependence on the neighboring threshold. As the neighboring threshold is lowered, the FWHM and the fraction of $cls=1$ events decrease, while the proportion of $cls=2$ events increases. The residual plateau width for $cls=1$ can be smaller than the SiPM pitch (2 mm), because muons traversing near the boundary between two SiPM channels tend to produce signals in two adjacent channels and are therefore categorized as $cls=2$.

In the experimental analysis, we use the cluster thresholds defined in Sec.~\ref{sec:cluster} and the good muon events selected in Sec.~\ref{sec:eff}. The spatial resolution of a DUT is evaluated from the residual distribution between the detected cluster position on the DUT and the extrapolated track reconstructed with the other layers. The measured residual variance, $\sigma_{r}^{2}$, includes contributions from both the intrinsic spatial resolution of the DUT, $\sigma_{d}$, and the resolution of the extrapolated track, $\sigma_{f}$:
\begin{equation}
\sigma_{r}^{2} = \sigma_{d}^{2} + \sigma_{f}^{2} \; .
\end{equation}
Since tracks are reconstructed with a straight-line fit, the extrapolated hit position is a linear combination of the reference layers. Accordingly, the resolution of the extrapolated position at module $m$ can be expressed as~\cite{Wang:2022TNS}:
\begin{equation}
\begin{aligned}
\sigma_{f,m}^{2} &= \sum_{j\neq m} w_{mj}^{2}\sigma_{d,j}^{2}, \\[3pt]
w_{mj} &= \frac{(z_j-\bar z)(z_m-\bar z)}{\sum_{j\ne m}(z_j-\bar z)^2} + \frac{1}{N-1}, \\[3pt]
\bar z &= \frac{1}{N-1}\sum_{j\ne m} z_j \; ,
\end{aligned}
\label{eq:pos}
\end{equation}
where $N$ is the total number of long modules, $z_j$ denotes the vertical position of module $j$ along the $z$ axis and $w_{mj}$ refers to the weight of each layer uncertainty extracted from the error progation of straight line fit formulae. The spatial resolution of each module was then obtained by solving the set of linear equations for all four modules.

\begin{figure}[htbp]
  \centering
  \begin{subfigure}[t]{0.32\textwidth}
    \centering
    \begin{overpic}[width=\linewidth]{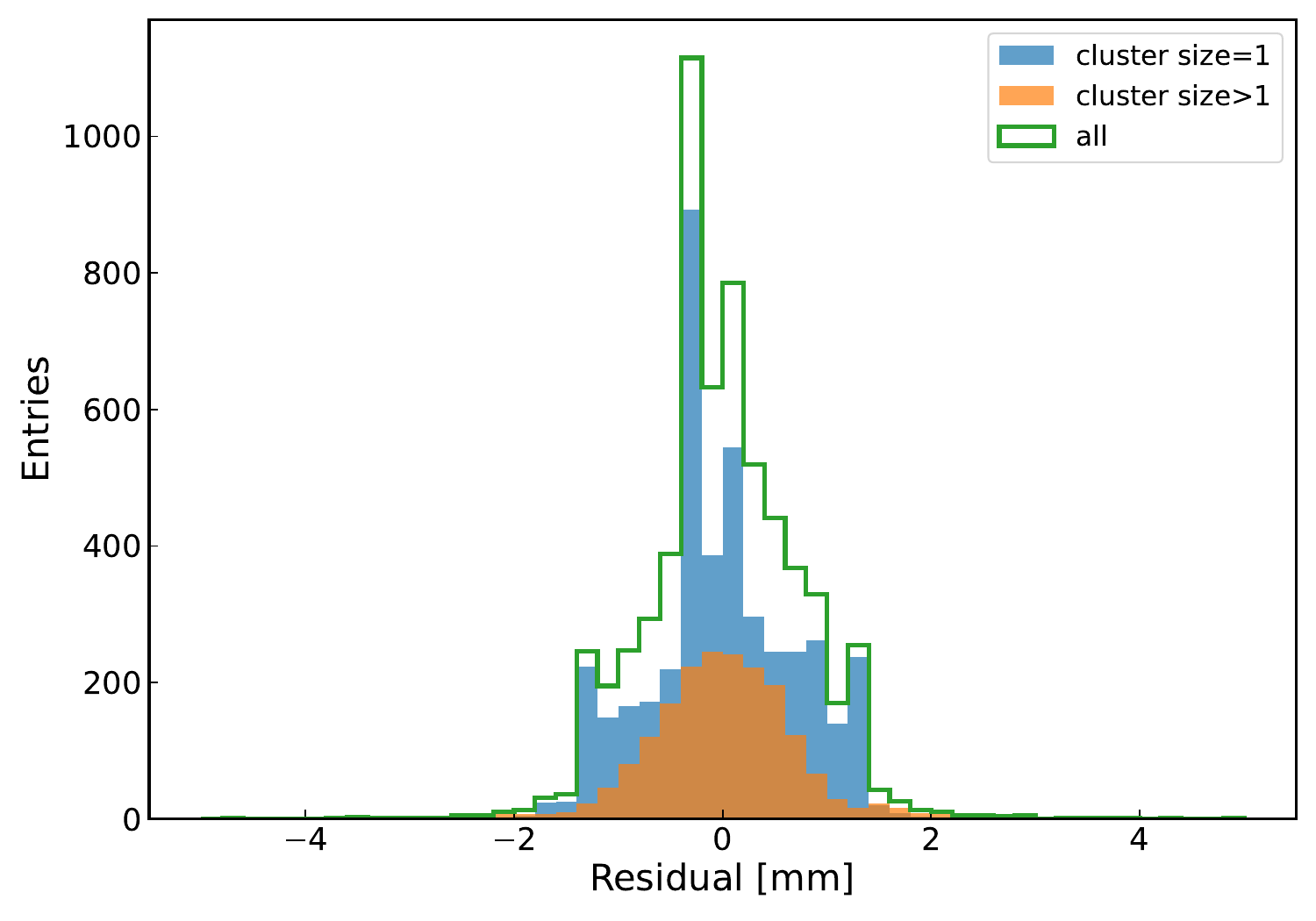} 
      \put(12,60){\small\textbf{(a)}}
    \end{overpic}
  \end{subfigure}
  \hfill
  \hspace{-0.2cm}
  \begin{subfigure}[t]{0.32\textwidth}
    \centering
    \begin{overpic}[width=\linewidth]{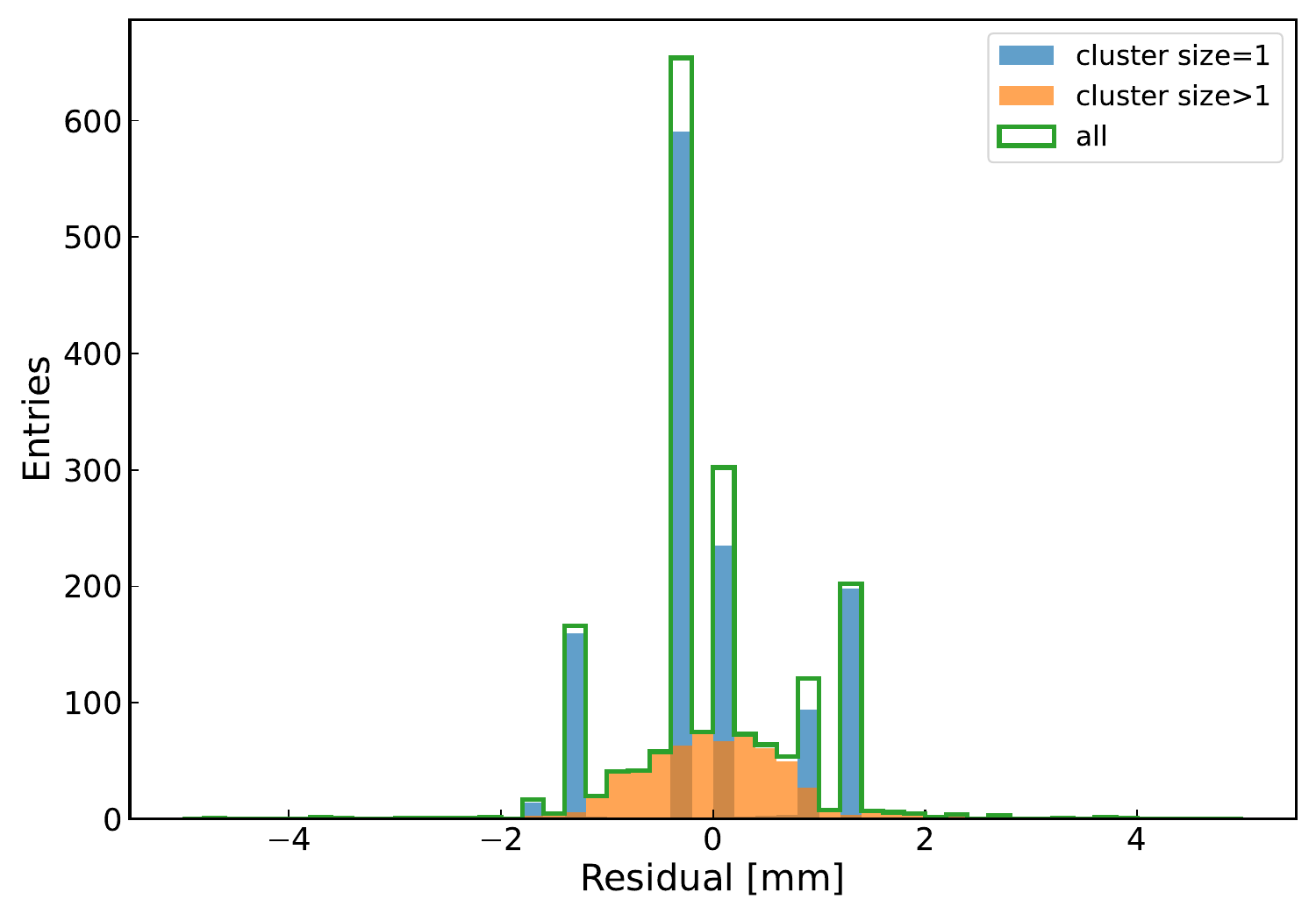}
      \put(12,60){\small\textbf{(b)}}
    \end{overpic}
  \end{subfigure}
  \hfill
  \hspace{-0.2cm}
  \begin{subfigure}[t]{0.32\textwidth}
    \centering
    \begin{overpic}[width=\linewidth]{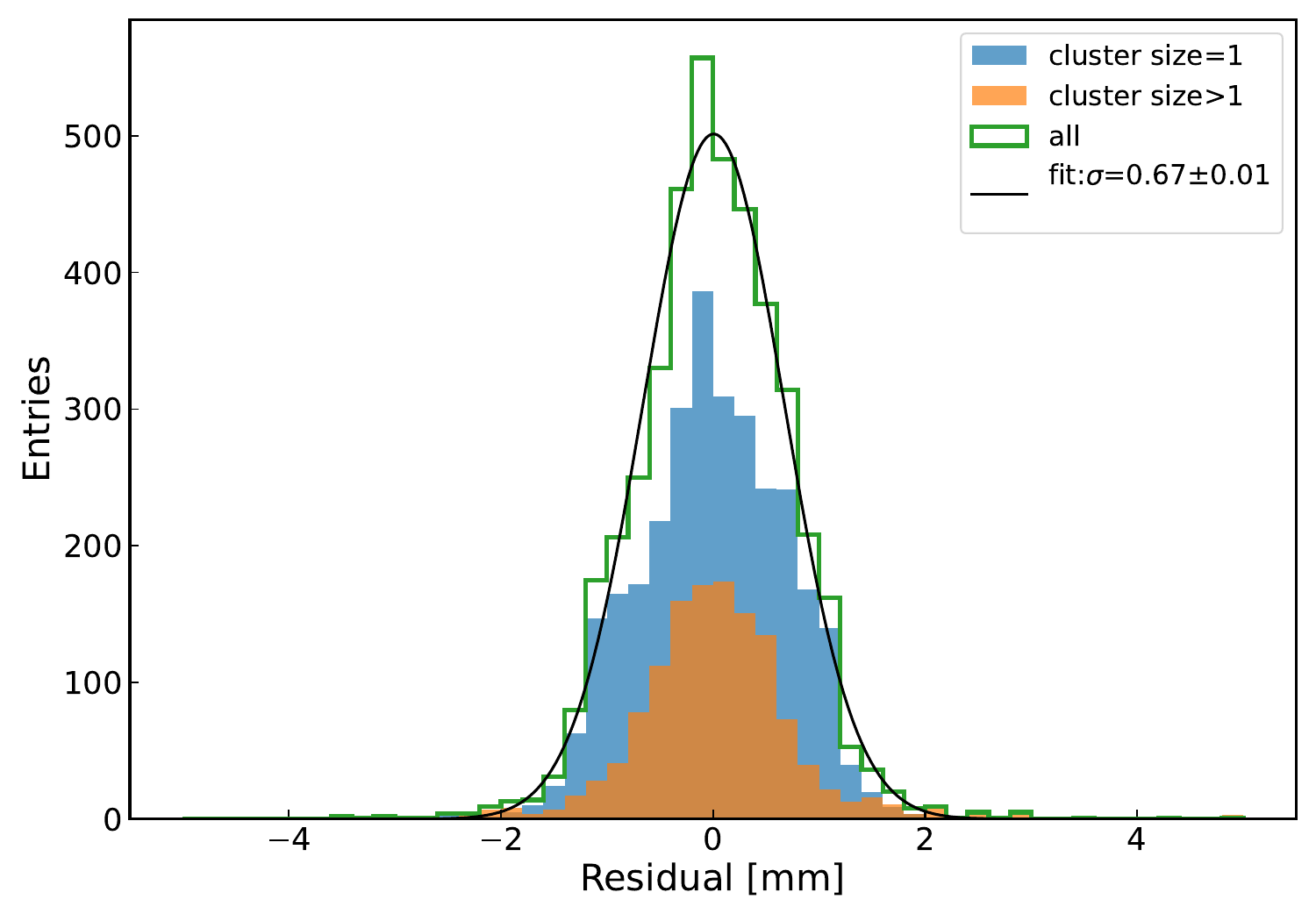}
      \put(12,60){\small\textbf{(c)}}
    \end{overpic}
  \end{subfigure}
  \caption{Residual distributions of the DUT in layer~2: (a) for all detected good events with clusters of $cls=1$ or $cls>1$ on the DUT, (b) for good events with extrapolated tracks reconstructed exclusively from $cls=1$ clusters in the three reference layers, and (c) for good events after excluding tracks reconstructed exclusively from $cls=1$ clusters in the three reference layers, together with a Gaussian fit of their residual distribution.}
  \label{fig:residual_cls}
\end{figure}

For the detected good events defined in Sec.~\ref{sec:eff}, the residual variance is usually determined by fitting the residual distribution with a Gaussian function. However, as shown in Fig.~\ref{fig:residual_cls}(a), the residual distribution of the DUT in layer~2 exhibits a Gaussian-like shape for clusters with $cls>1$, whereas discrete spikes are observed for clusters with $cls=1$.

Because the SiPM pitch is larger than the fiber diameter, about 68\% of muon clusters correspond to $cls=1$, as shown in Fig.~\ref{fig:clusters}. In these cases, the reconstructed cluster positions by the CoG method are fixed at the SiPM channel centers. As a result, roughly 30\% of tracks are reconstructed exclusively from $cls=1$ clusters in only three reference layers. Such tracks with hits on three SiPM centers lead to systematic reconstruction biases imposed by the fixed detector layout of the SiPMs as illustrated in Fig.~\ref{fig:trk_cls}. Furthermore, since only DUT clusters with residuals smaller than 5~mm are accepted, the residual distribution of such biased tracks shows discrete spikes for $cls=1$ clusters, as seen in Fig.~\ref{fig:residual_cls}(b). Similar patterns are also observed for other layers. 
\begin{figure}[!htbp]
\centering
\includegraphics[width=0.8\linewidth]{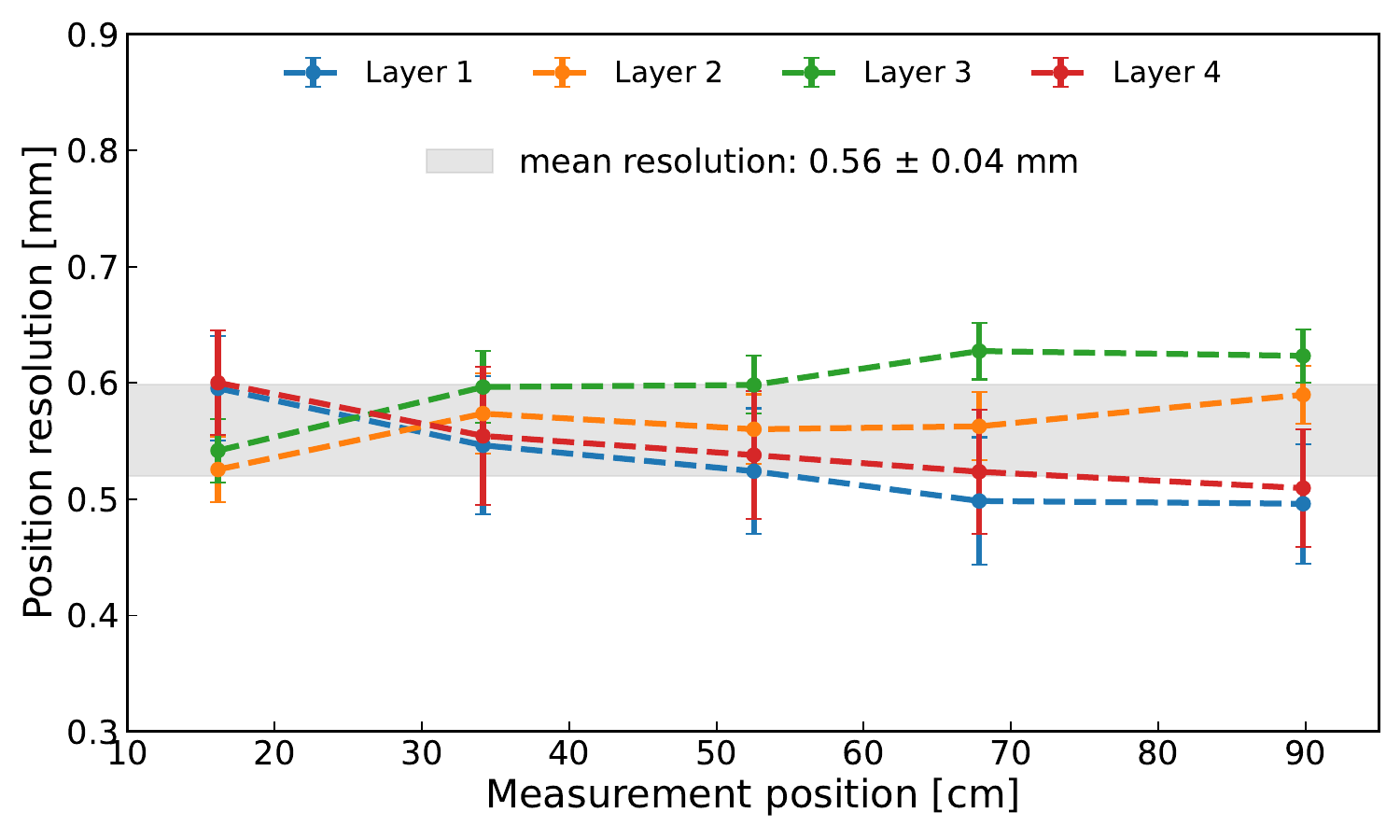}
\caption{Spatial resolution of the four 100-cm long SciFi modules measured with muons at different distances from the readout end. The error bars are obtained from the propagation of fitting uncertainties through the solution of the linear equations. The shaded band indicates the average spatial resolution across all layers and measurement positions.}
\label{fig:posRes}
\end{figure}

\begin{figure}[!htbp]
\centering
\includegraphics[width=0.8\linewidth]{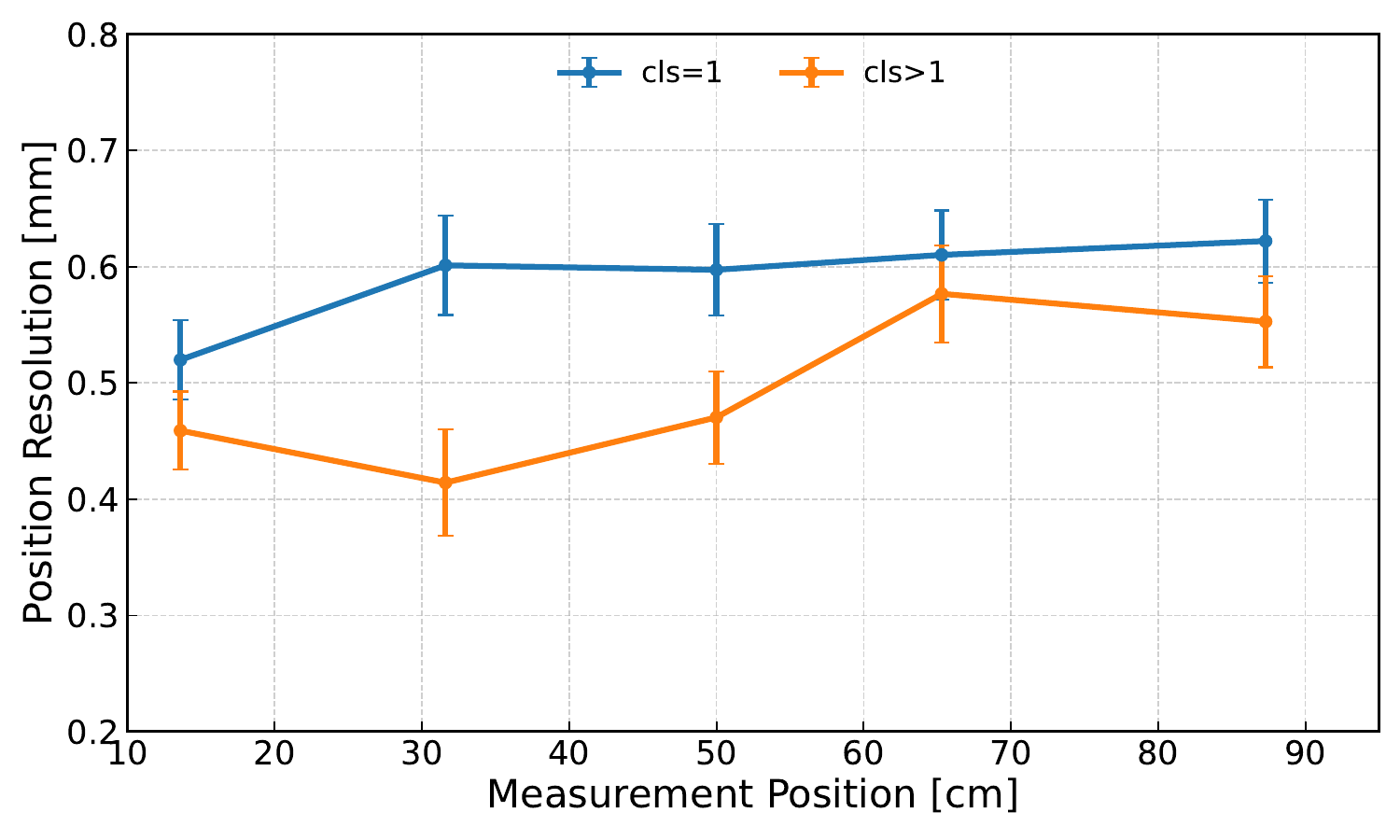}
\caption{Spatial resolution of events with $cls=1$ and $cls>1$ on layer~3. }
\label{fig:posResCls}
\end{figure}

After excluding tracks reconstructed solely from $cls=1$ clusters in the reference layers, the combined residual distribution of DUT clusters with $cls=1$ and $cls>1$ is well described by a Gaussian fit, from which $\sigma_{r}$ is extracted (Fig.~\ref{fig:residual_cls}(c)). By solving Eq.~\ref{eq:pos}, the intrinsic spatial resolutions including contributions from both $cls=1$ and $cls>1$ events were obtained for each module. Figure~\ref{fig:posRes} summarizes the spatial resolutions of all four SciFi modules at different measurement positions. The average resolution over all modules and positions is approximately $0.56 \pm 0.04$~mm and the modules exhibit an approximately uniform response along their length. As an illustration, Fig.~\ref{fig:posResCls} shows the intrinsic spatial resolutions for $cls=1$ and $cls>1$ events in layer 3, obtained using the same procedure with the combined position resolutions of the other layers fixed to the values extracted previously.

\section{Imaging experiment}\label{sec:6}
Cosmic-ray muons passing through materials of different atomic numbers experience distinct multiple scattering angles, which can be approximated by the formula:
\begin{equation}
\theta_{0} = \frac{13.6~\mathrm{MeV}}{\beta p} z
\sqrt{\frac{x}{X_{0}}}
\left[1 + 0.038\ln\left(\frac{x}{X_{0}}\right)\right],
\label{eq:MCS}
\end{equation}
where $\theta_{0}$ is the RMS of the projected multiple scattering angle, $p$ is the particle momentum in MeV/$c$, $\beta$ is the particle velocity relative to the speed of light, $z$ is the muon charge number, $x$ is the material thickness, and $X_{0}$ is the radiation length of the material. By measuring the deflection angles of muons before and after traversing a target, materials with different atomic numbers can thus be discriminated based on muon scattering.

\begin{figure}[!htbp]
\centering
\vspace{0cm}
\includegraphics[width=0.7\linewidth]{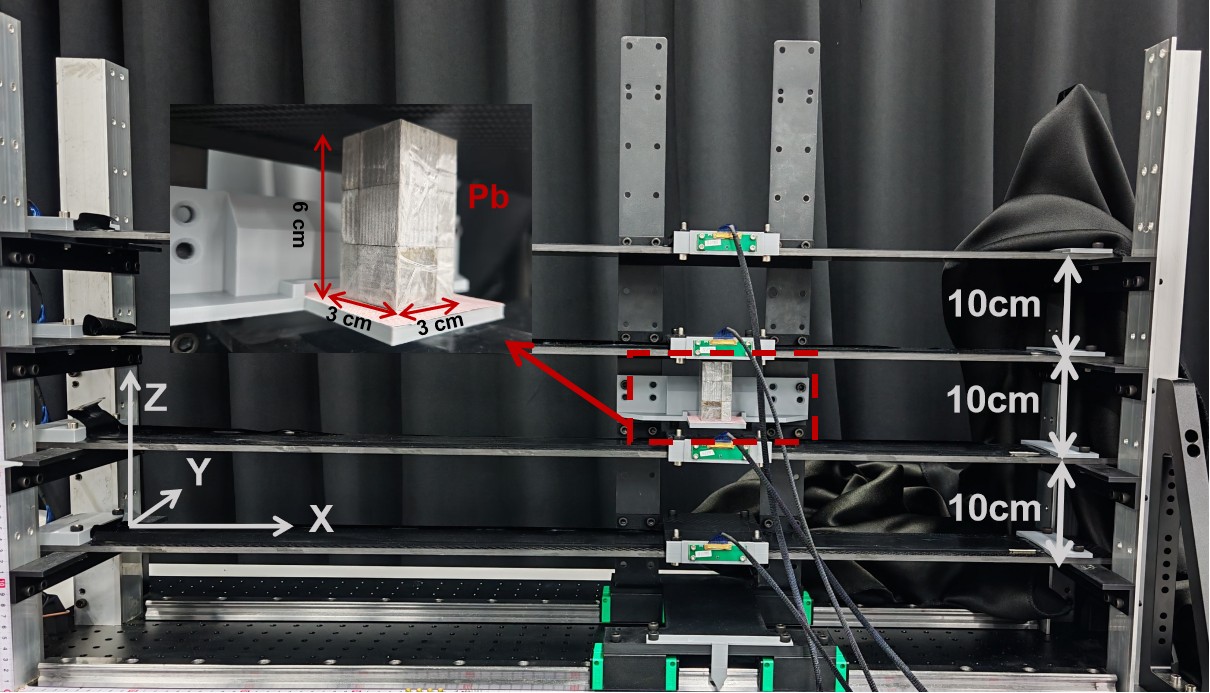}
\caption{Muon scattering tomography setup with a $\rm{3~cm\times3~cm\times6~cm}$ lead block placed between the upper and lower detectors.}
\label{fig:imageSetup}
\end{figure}

In the imaging setup shown in Fig.~\ref{fig:imageSetup}, the upper detector consists of layers~1, 2, 5, and 6, while the lower detector consists of layers~3, 4, 7, and 8. The separation $L$ between the two long modules in either the upper or lower detector is 10~cm. The upper and lower detectors are used to record muon trajectories before and after the target, respectively. The sensitive acceptance corresponding to a fiducial volume of approximately $\rm 5.5~cm \times 5.5~cm \times 30~cm$ is relatively small. Thus, the following imaging test serves mainly as a proof-of-principle demonstration of the capabilities of the developed modules for muon tomography.

As evaluated in Sec.~\ref{sec:5}, the mean position resolution $\sigma_{d}$ of a long SciFi module is approximately 0.56~mm. The corresponding angular resolution $\sigma_{\theta}$ of either the upper or lower detector is estimated as $\sigma_{\theta} = \sqrt{2}\sigma_{d}/L \approx 8~\mathrm{mrad}$. For a 4~GeV muon traversing a 6~cm thick lead block, the expected RMS scattering angle $\theta_{0}$ is about 12~mrad. Therefore, the imaging setup is expected to distinguish a $\rm{3~cm\times3~cm\times6~cm}$ lead block placed between the upper and lower detectors in air.

After about ten days of data acquisition in muon-test mode, roughly 4000 events containing valid tracks in both the upper and lower detectors were recorded. Point of Closest Approach (PoCA) ~\cite{schultz2003cosmic} points were reconstructed for events with deviation angles greater than 6~mrad. A $k$-Nearest Neighbors (k-NN) algorithm was subsequently applied to count the number of neighboring points within a defined distance around each PoCA point, and those with a small number of neighbors were removed as noise outliers. 

\begin{figure}[htbp]
  \centering
  \begin{subfigure}[t]{0.33\textwidth}
    \centering
    \begin{overpic}[width=\linewidth]{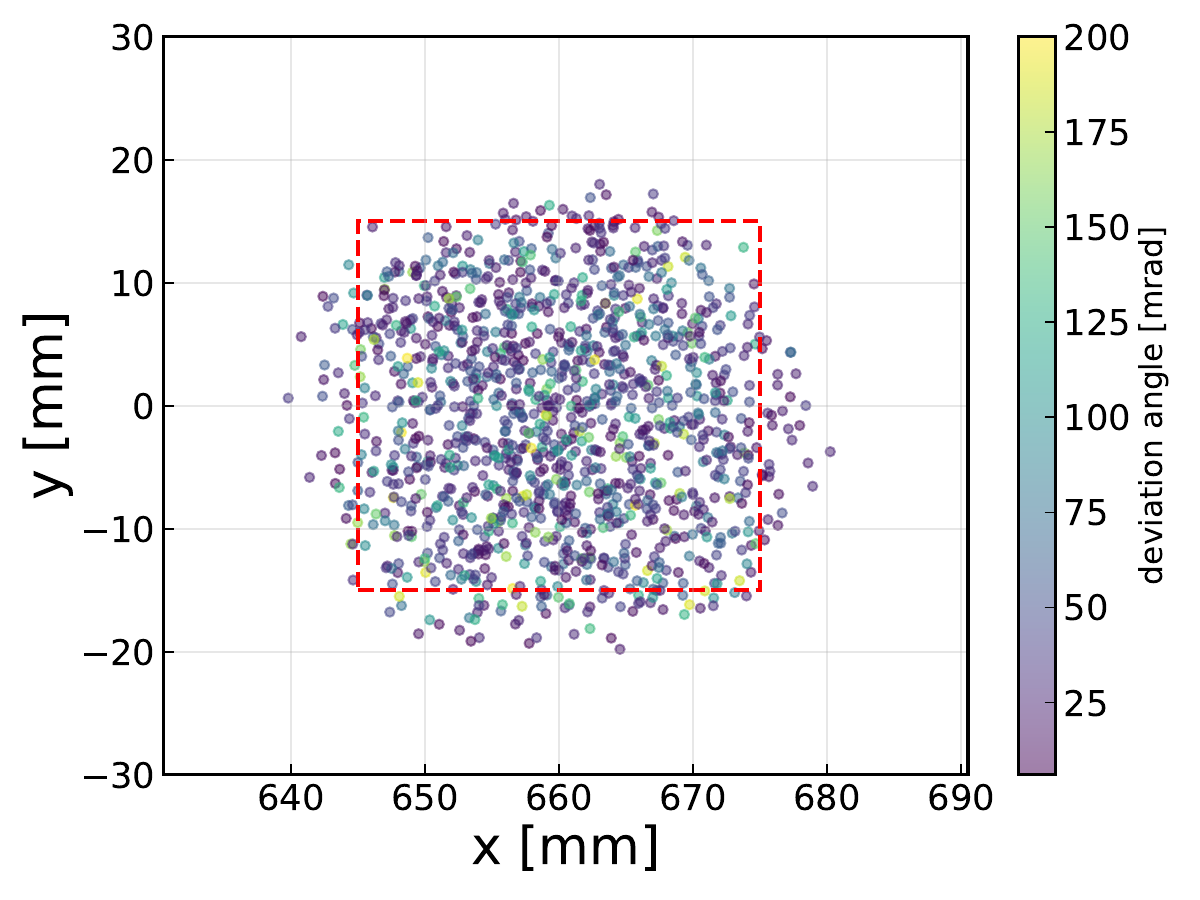} 
      \put(14,65){\small\textbf{(a)}}
    \end{overpic}
  \end{subfigure}
  \hfill
  \hspace{-0.2cm}
  \begin{subfigure}[t]{0.33\textwidth}
    \centering
    \begin{overpic}[width=\linewidth]{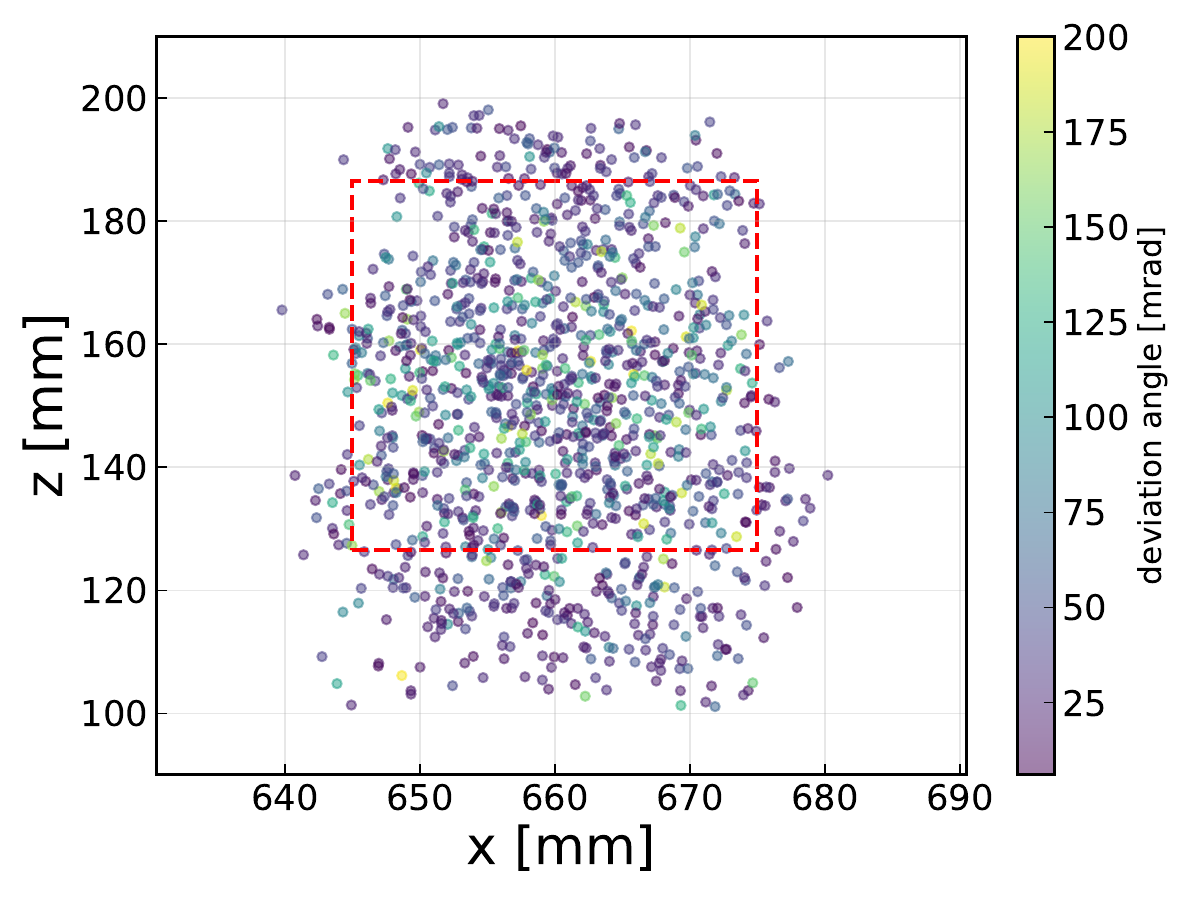}
      \put(14,65){\small\textbf{(b)}}
    \end{overpic}
  \end{subfigure}
  \hfill
  \hspace{-0.2cm}
  \begin{subfigure}[t]{0.33\textwidth}
    \centering
    \begin{overpic}[width=\linewidth]{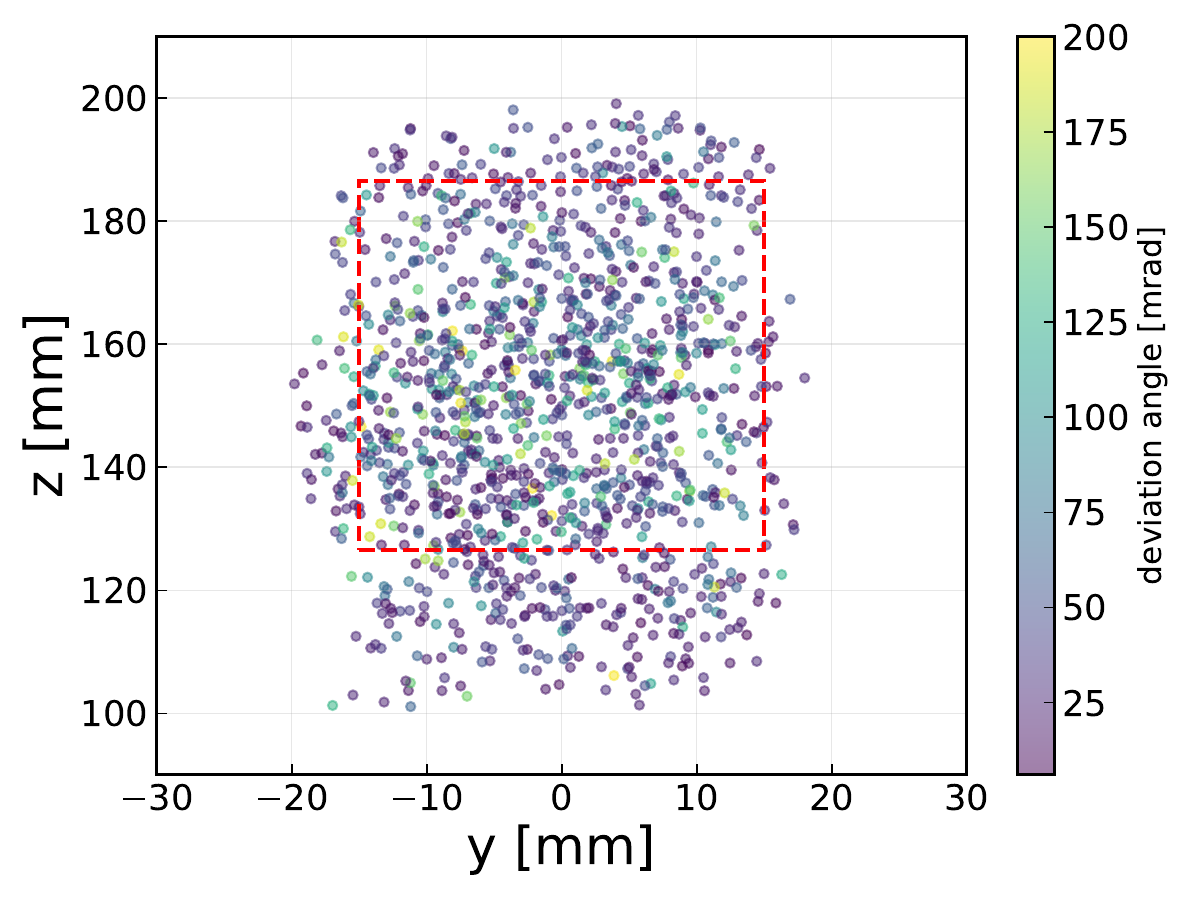}
      \put(14,65){\small\textbf{(c)}}
    \end{overpic}
  \end{subfigure}
  \caption{Reconstructed PoCA scattering maps projected onto the $\rm{XY}$ (a), $\rm{XZ}$ (b), and $\rm{YZ}$ (c) planes. The color of each point represents the corresponding event’s deviation angle.}
  \label{fig:imageRst}
\end{figure}

The reconstructed projections are presented in Fig.~\ref{fig:imageRst}, where color of each point represents the corresponding event’s deflection angle. Most PoCA points with large deviation angles cluster within the true lead block region. The $\rm{XY}$ projections clearly reveal the block region, the ambiguity of PoCA points along the $z$ axis in both the $\rm{XZ}$ and $\rm{YZ}$ projections mainly results from the limited geometrical acceptance of the imaging setup, where muons undergoing large deflections are more likely to fall outside the lower detector acceptance~\cite{luo2022development}. Overall, the results demonstrate the feasibility of the developed SciFi detector modules for compact muon tomography imaging applications.

\section{Conclusion}
\label{sec:Conclusion}
In this work, we designed and constructed four compact 100 cm long SciFi modules, each comprising two staggered layers of 1 mm diameter plastic scintillating fibers read out at one end by a one-dimensional SiPM array with a 2 mm pitch. The modules were characterized using cosmic-ray muons triggered by 10 cm short SciFi modules positioned at different distances along their length, with data acquired by Citiroc 1A–based electronics. After dedicated channel calibration, signal clustering, track reconstruction, and alignment, the modules achieved detection efficiencies above 97\% and an average spatial resolution of about 0.56~mm. Both parameters were found to be uniform across the full module length. Building upon these characterization results, a proof-of-principle imaging experiment was conducted using the SciFi modules arranged in a muon scattering tomography configuration. The reconstructed image obtained via the PoCA method revealed the profile of the target lead block, demonstrating the system’s capability for muon-based imaging.

Compared with previously developed scintillating fiber detectors for muon tomography, the present design is more compact and avoids the laborious fiber-grouping process, while providing sub-millimeter spatial resolution superior to that of bulk scintillator detectors. These features demonstrate the potential of the proposed modules for building compact, high-resolution muon tomography systems. Future work focusing on the muon tomography prototype with multiplexed SiPM readout is under development toward compact and cost-effective large-scale applications.

\section*{Acknowledgements}

This work was supported by the National Natural Science Foundation of China under Grant No. 12205174.

\bibliographystyle{elsarticle-num-names} 
\bibliography{refs}

@article{Navas:2024,
  author = {Navas, S. and Amsler, C. and Gutsche, T. and Hanhart, C. and Hernández-Rey, J.J. and Lourenço, C. and Masoni, A. and et al.},
  title = {Review of particle physics},
  journal = {Phys. Rev. D},
  volume = {110},
  number = {3},
  pages = {030001},
  year = {2024}
}

@article{Borozdin2003,
  title={Radiographic imaging with cosmic-ray muons},
  author={Borozdin, Konstantin N and Hogan, Gary E and Morris, Christopher and Priedhorsky, William C and Saunders, Alexander and Schultz, Larry J and Teasdale, Margaret E},
  journal={Nature},
  volume={422},
  number={6929},
  pages={277--277},
  year={2003},
  publisher={Nature Publishing Group UK London}
}

@article{Bonomi:2020,
  title={Applications of cosmic-ray muons},
  author={Bonomi, G and Checchia, P and D’Errico, M and Pagano, D and Saracino, G},
  journal={Progress in Particle and Nuclear Physics},
  volume={112},
  pages={103768},
  year={2020},
  publisher={Elsevier}
}

@article{Tanaka:2023,
  title={Muography},
  author={Tanaka, Hiroyuki KM and Bozza, Cristiano and Bross, Alan and Cantoni, Elena and Catalano, Osvaldo and Cerretto, Giancarlo and Giammanco, Andrea and Gluyas, Jon and Gnesi, Ivan and Holma, Marko and others},
  journal={Nature Reviews Methods Primers},
  volume={3},
  number={1},
  pages={88},
  year={2023},
  publisher={Nature Publishing Group UK London}
}

@article{Morishima:2017ghw,
  title={Discovery of a big void in Khufu’s Pyramid by observation of cosmic-ray muons},
  author={Morishima, Kunihiro and Kuno, Mitsuaki and Nishio, Akira and Kitagawa, Nobuko and Manabe, Yuta and Moto, Masaki and Takasaki, Fumihiko and Fujii, Hirofumi and Satoh, Kotaro and Kodama, Hideyo and others},
  journal={Nature},
  volume={552},
  number={7685},
  pages={386--390},
  year={2017},
  publisher={Nature Publishing Group UK London}
}

@article{Liu:2024,
  title={Deep investigation of muography in discovering geological structures in mineral exploration: a case study of Zaozigou gold mine},
  author={Liu, Guorui and Yao, Kaiqiang and Niu, Feiyun and Li, Zhuodai and Tian, Heng and Li, Jiangkun and Luo, Xujia and Jin, Long and Gao, Jinlei and Rong, Jian and others},
  journal={Geophysical Journal International},
  volume={237},
  number={1},
  pages={588--603},
  year={2024},
  publisher={Oxford University Press}
}

@article{Mahon:2018,
  title={First-of-a-kind muography for nuclear waste characterization},
  author={Mahon, David and Clarkson, Anthony and Gardner, Simon and Ireland, David and Jebali, Ramsey and Kaiser, Ralf and Ryan, Matthew and Shearer, Craig and Yang, Guangliang},
  journal={Philosophical Transactions of the Royal Society A},
  volume={377},
  number={2137},
  pages={20180048},
  year={2019},
  publisher={The Royal Society Publishing}
}

@article{Barnes:2023,
  title={Cosmic-ray tomography for border security},
  author={Barnes, Sarah and Georgadze, Anzori and Giammanco, Andrea and Kiisk, Madis and Kudryavtsev, Vitaly A and Lagrange, Maxime and Pinto, Olin Lyod},
  journal={Instruments},
  volume={7},
  number={1},
  pages={13},
  year={2023},
  publisher={MDPI}
}

@article{Lesparre:2012,
  title={Design and operation of a field telescope for cosmic ray geophysical tomography},
  author={Lesparre, N and Marteau, J and D{\'e}clais, Y and Gibert, Dominique and Carlus, B and Nicollin, Florence and Kergosien, Bruno},
  journal={Geoscientific Instrumentation, Methods and Data Systems},
  volume={1},
  number={1},
  pages={33--42},
  year={2012},
  publisher={Copernicus GmbH}
}

@article{Anstasio:2013,
  title={The MU-RAY detector for muon radiography of volcanoes},
  author={Anastasio, A and Ambrosino, Fabio and Basta, D and Bonechi, L and Brianzi, M and Bross, A and Callier, S and Caputo, A and Ciaranfi, R and Cimmino, Luigi and others},
  journal={Nuclear Instruments and Methods in Physics Research Section A: Accelerators, Spectrometers, Detectors and Associated Equipment},
  volume={732},
  pages={423--426},
  year={2013},
  publisher={Elsevier}
}

@article{Wang:2024,
  title={Electronics design for a muon imaging system using triangular plastic scintillators with WLS fiber readouts},
  author={Wang, ZY and Wang, YG and Li, X and Zhao, YX and Liang, YT and Liang, Z and Zhang, YS and Tang, ZB and Li, C},
  journal={Journal of Instrumentation},
  volume={19},
  number={02},
  pages={P02033},
  year={2024},
  publisher={IOP Publishing}
}

@article{Yu:2025,
  title={MuGrid-v2: A novel scintillator detector for multidisciplinary applications},
  author={Yu, Tao and Ning, Yunsong and Yuan, Yi and Zhao, Shihan and Qi, Songran and Sun, Mingchen and Li, Yuye and Liu, Zhirui and Bai, Aiyu and Liu, Hesheng and others},
  journal={Journal of Applied Physics},
  volume={138},
  number={2},
  year={2025},
  publisher={AIP Publishing}
}

@article{Clarkson:2014,
  title={The design and performance of a scintillating-fibre tracker for the cosmic-ray muon tomography of legacy nuclear waste containers},
  author={Clarkson, Anthony and Hamilton, David J and Hoek, Matthias and Ireland, David G and Johnstone, JR and Kaiser, Ralf and Keri, Tibor and Lumsden, Scott and Mahon, David F and McKinnon, Bryan and others},
  journal={Nuclear Instruments and Methods in Physics Research Section A: Accelerators, Spectrometers, Detectors and Associated Equipment},
  volume={745},
  pages={138--149},
  year={2014},
  publisher={Elsevier}
}

@article{Anbarjafari:2021,
  title={Atmospheric ray tomography for low-Z materials: implementing new methods on a proof-of-concept tomograph},
  author={Anbarjafari, Gholamreza and Anier, Aivo and Avots, Egils and Georgadze, Anzori and Hektor, Andi and Kiisk, Madis and Kutateladze, Marius and Lepp, T{\~o}nu and M{\"a}gi, M{\"a}rt and Pastsuk, Vitali and others},
  journal={arXiv preprint arXiv:2102.12542},
  year={2021}
}

@article{Chen:2023,
  title={Towards a muon scattering tomography system for both low-Z and high-Z materials},
  author={Chen, Jiahui and Li, Huiling and Li, Yiyue and Liu, Pingcheng},
  journal={Journal of Instrumentation},
  volume={18},
  number={08},
  pages={P08008},
  year={2023},
  publisher={IOP Publishing}
}

@article{lv2025multiplexed,
  title={Multiplexed SiPM Readout of Plastic Scintillating Fiber Detector for Muon Tomography},
  author={Lv, Chenghan and Hu, Kun and Li, Huiling and Liang, Hui and Liu, Cong and Wang, Hongbo and Wu, Zibing and Xu, Weiwei},
  journal={arXiv preprint arXiv:2511.16185},
  year={2025}
}

@misc{kurarayScintillatingFibers,
  author       = "{Kuraray Co., Ltd.}",
  title        = "{Plastic Scintillating Fibers}",
  howpublished = "\url{http://kuraraypsf.jp/psf/sf.html}",
  year = 2025,
  note         = "Accessed: 2025-09-01"
}

@misc{HamamatsuS13360-2050VE,
  author       = "{Hamamatsu Photonics K.K.}",
  title        = "{MPPC S13360-2050VE (TSV chip-on-board package)}",
  howpublished = "\url{https://www.hamamatsu.com/eu/en/product/optical-sensors/mppc/mppc_mppc-array/S13360-2050VE.html}",
  year = 2024,
  note         = "Accessed: 2025-09-01"
}

@misc{Citiroc1A,
  author = "{Weeroc}",
  title  = "{Citiroc 1A -- Scientific instrumentation SiPM readout chip}",
  author = "{Weeroc}",
  howpublished = "Available at: \url{https://www.weeroc.com/read_out_chips/citiroc-1a}",
  year = 2019,
  note = "Accessed: 2025-09-01"
}

@article{Impiombato:2020,
  title={Use of the Peak-Detector mode for gain calibration of SiPM sensors with ASIC CITIROC read-out},
  author={Impiombato, Domenico and Segreto, Alberto and Catalano, Osvaldo and Giarrusso, Salvatore and Mineo, Teresa},
  journal={Journal of Instrumentation},
  volume={15},
  number={04},
  pages={C04007},
  year={2020},
  publisher={IOP Publishing}
}

@article{Wu:2024,
  title={Multi-channel readout electronics of silicon photomultipliers for plastic scintillating fiber detector},
  author={Wu, Zibing and Hu, Kun and Li, Huiling and Ren, Xiangxiang and Wang, Hongbo and Xu, Weiwei},
  journal={Journal of Instrumentation},
  volume={19},
  number={11},
  pages={C11014},
  year={2024},
  publisher={IOP Publishing}
}

@misc{Blobel2007,
  author = {Blobel, V.},
  title = {Alignment of track detectors -- Large scale optimization},
  howpublished = {\url{https://www.desy.de/~sschmitt/blobel/desy07talk.pdf}},
  year = 2007,
  note = {DESY Seminar on Computing in High Energy Physics, Hamburg, Germany}
}

@article{GEANT4:2002zbu,
  title={Geant4—a simulation toolkit},
  author={Agostinelli, Sea and Allison, John and Amako, K al and Apostolakis, John and Araujo, Henrique and Arce, Pedro and Asai, Makoto and Axen, D and Banerjee, Swagato and Barrand, GJNI and others},
  journal={Nuclear instruments and methods in physics research section A: Accelerators, Spectrometers, Detectors and Associated Equipment},
  volume={506},
  number={3},
  pages={250--303},
  year={2003},
  publisher={Elsevier}
}

@article{guan2015parametrization,
  title={A parametrization of the cosmic-ray muon flux at sea-level},
  author={Guan, Mengyun and Chu, Ming-Chung and Cao, Jun and Luk, Kam-Biu and Yang, Changgen},
  journal={arXiv preprint arXiv:1509.06176},
  year={2015}
}

@article{Wang:2022TNS,
  title={A high spatial resolution muon tomography prototype system based on micromegas detector},
  author={Wang, Yu and Zhang, Zhiyong and Liu, Shubin and Shen, Zhongtao and Feng, Changqing and Liu, Jianguo and Liu, Yulin},
  journal={IEEE Transactions on Nuclear Science},
  volume={69},
  number={1},
  pages={78--85},
  year={2021},
  publisher={IEEE}
}

@book{schultz2003cosmic,
  title={Cosmic ray muon radiography},
  author={Schultz, Larry Joe},
  year={2003},
  publisher={Portland State University}
}

@article{luo2022development,
  title={Development and commissioning of a compact Cosmic Ray Muon imaging prototype},
  author={Luo, Xujia and Wang, Quanxiao and Qin, Kemian and Tian, Heng and Fu, Zhiqiang and Zhao, Yanwei and Shen, Zhongtao and Liu, Hao and Fu, Yuanyong and Liu, Guorui and others},
  journal={Nuclear Instruments and Methods in Physics Research Section A: Accelerators, Spectrometers, Detectors and Associated Equipment},
  volume={1033},
  pages={166720},
  year={2022},
  publisher={Elsevier}
}

\end{document}